\documentclass[sigconf]{acmart}
\usepackage{multirow}
\usepackage{array}
\usepackage{graphicx}
\usepackage{algorithm}
\usepackage{algorithmic}
\usepackage{xspace}
\usepackage{tabularx}
\usepackage{balance}
\usepackage{subfig}

\graphicspath{{figs/}}

\newcommand{\ours}{SUI-Attack\xspace}

\newcommand{\hide}[1]{}

\copyrightyear{2023}
\acmYear{2023}
\setcopyright{acmlicensed}\acmConference[CIKM '23]{Proceedings of the 32nd ACM International Conference on Information and Knowledge Management}{October 21--25, 2023}{Birmingham, United Kingdom}
\acmBooktitle{Proceedings of the 32nd ACM International Conference on Information and Knowledge Management (CIKM '23), October 21--25, 2023, Birmingham, United Kingdom}
\acmPrice{15.00}
\acmDOI{10.1145/3583780.3615062}
\acmISBN{979-8-4007-0124-5/23/10}

\begin{document}

\title{Single-User Injection for Invisible Shilling Attack against Recommender Systems}

\author{Chengzhi Huang}
\orcid{0009-0009-4194-0393}
\affiliation{%
  \institution{Key Laboratory of Multimedia Trusted Perception and Efficient Computing, Ministry of Education of China\\Xiamen University}
  \city{Xiamen}
  \country{China}
}
\email{edisonchen@stu.xmu.edu.cn}

\author{Hui Li}
\authornote{Corresponding author.}
\orcid{0000-0001-9139-3855}
\affiliation{%
  \institution{Key Laboratory of Multimedia Trusted Perception and Efficient Computing, Ministry of Education of China\\Xiamen University}
  \city{Xiamen}
  \country{China}
}
\email{hui@xmu.edu.cn}

\begin{abstract}

Recommendation systems (RS) are crucial for alleviating the information overload problem. Due to its pivotal role in guiding users to make decisions, unscrupulous parties are lured to launch attacks against RS to affect the decisions of normal users and gain illegal profits. Among various types of attacks, shilling attack is one of the most subsistent and profitable attacks. In shilling attack, an adversarial party injects a number of well-designed fake user profiles into the system to mislead RS so that the attack goal can be achieved. Although existing shilling attack methods have achieved promising results, they all adopt the attack paradigm of multi-user injection, where some fake user profiles are required. This paper provides the first study of shilling attack in an extremely limited scenario: only one fake user profile is injected into the victim RS to launch shilling attacks (i.e., single-user injection). We propose a novel single-user injection method SUI-Attack for invisible shilling attack. SUI-Attack is a graph based attack method that models shilling attack as a node generation task over the user-item bipartite graph of the victim RS, and it constructs the fake user profile by generating user features and edges that link the fake user to items. Extensive experiments demonstrate that SUI-Attack can achieve promising attack results in single-user injection. In addition to its attack power, SUI-Attack increases the stealthiness of shilling attack and reduces the risk of being detected. We provide our implementation at: \url{https://github.com/KDEGroup/SUI-Attack}.

\end{abstract}

\begin{CCSXML}
<ccs2012>
   <concept>
       <concept_id>10002978.10003022.10003026</concept_id>
       <concept_desc>Security and privacy~Web application security</concept_desc>
       <concept_significance>500</concept_significance>
       </concept>
   <concept>
       <concept_id>10002951.10003317.10003347.10003350</concept_id>
       <concept_desc>Information systems~Recommender systems</concept_desc>
       <concept_significance>500</concept_significance>
       </concept>
 </ccs2012>
\end{CCSXML}

\ccsdesc[500]{Security and privacy~Web application security}
\ccsdesc[500]{Information systems~Recommender systems}

\keywords{Shilling Attack, Recommender System, Adversarial Attack}

\maketitle

%!TEX root = main.tex

\section{Introduction}
\label{sec: intro}
With the rapid development of information technology, we are facing a huge growth of available information, causing the information overload problem~\cite{BorchersHR98}: it is difficult to effectively make decisions when facing too much information. 
Recommender systems (RS) are an essential tool to alleviate information overload and have been widely deployed in e-commerce platforms (e.g., Amazon and Taobao) and content-providing platforms (e.g., TikTok and YouTube), bringing massive revenue~\cite{WuHWZW23}.

However, the prevalence of RS has also attracted unscrupulous parties~\cite{DeldjooNM21}.
They try to attack RS to can gain illegal profits.
Among various attack types, shilling attack is one of the most subsistent and profitable attacks against RS~\cite{revisit}.
In shilling attack, an adversarial party injects a number of well-designed fake user profiles into the system to mislead RS so that the attack goal can be achieved~\cite{attack_survey, si2020shilling, DeldjooNM21}.
One main attack goal is to promote a target item: increase the possibility that the target item can be viewed/bought by people.
Studying how to spoof RS has become a hot direction in the RS community as it gives insights into improving the defense against malicious attacks~\cite{Zeng0JCL23}.

Much effort has been devoted to designing shilling attack methods.
Pioneering works (e.g., Random Attack~\cite{BurkeMBW05}, Bandwagon Attack~\cite{Burke2005Limited} and Segment Attack~\cite{BurkeMBW05}) mainly adopt heuristics~\cite{attack_survey}.
Recently, based on the idea of adversarial attack~\cite{YuanHZL19}, a great number of shilling attack approaches have sprung up, including but not limited to optimization based methods~\cite{revisit}, GAN based methods~\cite{aush,legup}, reinforcement learning based methods~\cite{SongLHWLLG20}, knowledge distillation based methods~\cite{YueHZM21} and pre-training based methods~\cite{Zeng0JCL23}.
Existing methods all adopt the same attack paradigm: inject some fake user profiles into the victim RS. 
We name such an attack paradigm \textit{multi-user injection}.
As more injected fake user profiles typically improve the attack performance but increase the risk of being detected, the number of the injected fake user profiles is typically not large, e.g., 50.

Although existing shilling attack methods have achieved promising attack performance~\cite{DeldjooNM21}, they all assume there is a trade-off between the number of fake users and the performance of the attack.
To our best knowledge, no work has studied and answered a critical question about shilling attack: How many fake user profiles are required to launch a successful shilling attack? 
In this paper, we study shilling attack in an extremely limited scenario: only one fake user profile is injected into the victim RS to launch shilling attacks (i.e., \textit{single-user injection}).
Adversarial attacks against different AI models in the extremely restricted settings (e.g., one-pixel attack in image classification~\cite{su2019one} and single-node attack against Graph Neural Networks~\cite{finkelshtein2022single,single}) have attracted considerable attention since they unveil the severe vulnerability of AI models: adversary can hoax the model with minimum effort.
Single-user attacks against RS, if possible, will result in the virtually undetectable attack as it is extremely difficult to identify the only fake user from plenty of real users.

In this paper, we propose a novel \underline{s}ingle-\underline{u}ser \underline{i}njection method for invisible shilling \underline{attack} against RS (i.e., \ours).
\ours is a graph based attack method that models shilling attack as a node generation task over the user-item bipartite graph of the victim RS.
\ours contains two phases: feature generation and edge generation. 
The feature generation phase aims to produce toxic fake user features that can guide the generation of edges that are connected to the fake user.
The edge generation phase connects the fake user to items in the user-item bipartite graph to ensure the injected fake user can affect the victim RS, and it is equivalent to filling the fake user profile with interaction history in contemporary shilling attack approaches.
The contributions of this work can be summarized as follows:
\begin{itemize}

\item We propose the idea of single-user injection. As far as we know, this is the first work to study shilling attacks in the extremely restricted scenario.

\item We design a novel single-user injection method \ours, increasing the stealthiness of shilling attacks and reducing the risk of being detected. \ours models the single-user injection task over the user-item bipartite graph and constructs the fake user profile by generating its user features and edges that link the fake user to items. 

\item We conduct extensive experiments to demonstrate that \ours can achieve promising attack results in single-user injection. In other words, shilling attacks against RS with single-user injection is achievable. Furthermore, in the traditional multi-user injection setting, \ours is shown to be effective and can cause comparable attack results compared to existing shilling attack methods, showing its flexibility in shilling attacks.

\end{itemize}

The remaining parts of this paper are organized as follows:
Section~\ref{sec:background} provides the background knowledge of this work. 
We describes the details of \ours in Section~\ref{sec:method}. 
Experimental results and analysis are presented in Section~\ref{sec:exp}. 
Section~\ref{sec:related_work} introduces the related work of this study. 
Section~\ref{sec:con} concludes this work.

%!TEX root = main.tex

\section{Background}
\label{sec:background}

In this section, we first provide some background knowledge of shilling attack and the extremely restricted setting of shilling attack (i.e., single-user injection) that we consider in this paper.

\vspace{5pt}
\noindent\textbf{Attack Goal:}
There are two types of shilling attack: promotion attack and nuke attack~\cite{attack_survey, si2020shilling}. 
Through injecting several fake user profiles into the target RS (i.e., the \textit{victim RS}), promotion attack aims to improve the ranking of the \textit{target item} in a user's recommendation list. 
The goal of nuke attack is opposite to promotion attack and it can be easily achieved by reversing the goal of the promotion attack. 
Hence, for simplicity, we focus on the promotion attack in this paper.

After a successful shilling attack, the target item should appear in as many users' recommendation lists as possible while the overall recommendation performance of the system is not affected~\cite{li2016data}. 
In addition to the traditional settings of shilling attack, in this paper, we impose an extremely restricted constraint on shilling attack and study the single-user-injection shilling attack: the attacker only inject one fake user profile to spoof RS.

\vspace{5pt}
\noindent\textbf{Attack Knowledge:}
We consider the most common setting of attack knowledge used by the existing studies of shilling attack~\cite{christakopoulou2019adversarial,aush}. 
The attackers do not have prior knowledge of the model architecture of the victim RS.
They cannot access the parameters of the victim RS model as well as the gradients during training.
However, attackers can access the most basic user-item historical data of the victim RS, i.e., user-item ratings.
User-item ratings in many RS (e.g., Amazon) are publicly accessible and can be crawled by attackers.

\vspace{5pt}
\noindent\textbf{Attack Capabilities:}
Typically, a more powerful shilling attack requires injecting more fake user profiles, making the attack more perceptible to the system.
Therefore, the number of fake user profiles and the number of maximum interacted items in each fake user profile are limited to $b_{user}$ and $b_{item}$ (i.e., the budget), respectively.
Note that, in the setting of single-user injection studied in this paper, $b_{user}$ is set to $1$.
However, we still report the case when $b_{user} > 1$ so that \ours can be compared to other shilling attack methods in the traditional multi-user injection setting.

%!TEX root = main.tex
\section{Our Method \ours}
\label{sec:method}

In this section, we illustrate the details of our proposed \ours.

\subsection{Overview}

We model the single-user-injection attack as a node generation process over the user-item bipartite graph.
The target is to generate a fake user node that can be used to guide the construction of the fake user profile for injection.
The user-item bipartite graph, where each edge between a user node and an item node indicates an historical user-item interaction and edge weights denote interaction features like ratings, is commonly used to model RS. 

\ours uses two phases, feature generation and edge generation, to generate the fake user profile, including its user features and edges connecting the fake user and items on the bipartite graph, for single-user-injection attack.

\subsection{Feature Generation}
User features (e.g., statistics of historical ratings), which are typically used to initialize the embedding layer in RS models, are an important source for RS to model user preferences. 
Attackers can also leverage user features to guide the construction of the fake user profile. 
However, unlike real users, the fake user does not have features as it does not have real interaction history. 
Feature generation phase aims to generate fake user's node features with strong toxicity that can guide the subsequent edge generation phase to generate destructive user-item interactions for the fake user to hoax RS.
Note that \textit{user features in fake user profile construction are not the same as user features modeled by the victim RS.}
The latter are unacquirable for attackers as they cannot access the details of the victim RS model.

\subsubsection{Selection of Node Features}
As different RS may have their own designed user/item features, we choose to adopt 10 prevalent RS features~\cite{williams2007defending, morid2014defending} that rely on the intrinsic information of the user-item bipartite graph and do not require specific user or item information (e.g., user demographics and item descriptions).
This way, \ours is not limited to specific feature designs and can be applied to different RS.
The definitions of the ten chosen features are as follows:
\begin{enumerate}

    \item \textbf{Rating Deviation from Mean Agreement (RDMA)} measures the average deviation of a user's ratings from the mean agreement for a set of target items:
    \begin{equation}
    \text{RDMA}_u=\frac{\sum_{i\in N(u)}\frac{|r_{u,i}-\bar{r}_i|}{M_i}}{|N_u|},
    \end{equation}
    where $N_u$ is the items that user $u$ has rated, $|N_u|$ is the number of items in $N_u$ (i.e., profile size), $M_i$ is the number of ratings received by the item $i$, $r_{u,i}$ denotes the ratings given by user $u$ on item $i$, and $\bar{r}_i$ denotes the mean rating of item $i$. The reciprocal of the number of ratings for each item ($M_i$) is used as a weight since items with more ratings are more likely to be rated accurately and the weights of their deviations should be reduced.

% \textbf{Rating Deviation from Mean Agreement(RDMA):} RDMA is a metric for detecting attackers in recommender systems. It measures the average deviation of a user's ratings from the mean agreement for a set of target items. The inverse of the numer of ratings for each item is used as a weight to account for the fact that items with more ratings are more likely to be rated accurately, and RDMA is calculated as follows:
% $$
% RDMA_u=\frac{\sum^{N_u}_{i=0}\frac{|r_{u,i}-\bar {r_i}|}{NR_i}}{N_u}
% $$
% where $N_u$ is the number of ratings that user $u$ has rated and $NR_i$ is the number of ratings provided for item $i$. And $r_{u,i}$ denotes the ratings given by user $u$ to item $i$, and $\bar {r_i}$ denote the mean rating of item $i$ across all users.

    \item \textbf{Length Variance (LengthVar):} LengthVar measures the variance of the number of interactions in a user's profile (i.e., profile size) and it is defined as follows:
    \begin{equation}
    \text{LengthVar}_u=\frac{|N_u|-\bar{|N|}}{\sum_{j\in U}(|N_j|-|N_u|)^2},
    \end{equation}
    where $U$ indicates the user set in RS and $\bar{|N|}$ is the average profile size in RS.

% \textbf{Length Variance(LengthVar):} LengthVar measures the variance of the number of ratings in a user's profile. Attacker profiles typically have a higher variance in the number of ratings than genuine profiles, as attackers are more likely to create profiles with a small number of ratings or a large number of ratings. LengthVar is calculated as follows:
% $$
% LengthVar_u=\frac{n_u-\bar n}{\sum_{k\in U}(n_k-n)^2}
% $$
% where $n_u$ is the total number of ratings in the system for user u. $U$ is the total number of users in the system. $\bar n$ is the average length of a profile in the system.

    \item \textbf{Filler Mean Variance (FMV):} FMV measures the deviation of a user's rating in a hypothesized filler partition from the mean rating for each item. The hypothesized filler partition contains randomly sampled items. We sample at most 50 items for each user profile as the hypothesized filler partition. Then, FMV is defined as:
    \begin{equation}
        \text{FMV}_u=\frac{1}{\left| H_{u} \right|} \underset{i\in H_{u}}{\sum}(r_{u,i}-\bar{r}_i)^2,
    \end{equation}
    where $H_{u}$ is the hypothesized filler partition in the user profile of $u$ and $\left| H_{u} \right|$ indicates the number of items in $H_{u}$.

% \textbf{Filler Mean Variance(FMV):} FMV measures the variance of a user's rating in a hypothesized filler partition from the mean rating for each of the items. The hypothesized filler partition is a set of items that are likely to be rated by attackers in an attempt to increase the size of their user profiles. FMV is calculated as follow:
% $$
% FMV_u=\frac {1}{U_{u^{F_m}}} \underset{i\in U_{u^{F_m}}}{\sum}(r_{u,i}-\bar {r_i})^2
% $$
% where $U_{u^{F_m}}$ is the partition of the profile of user $u$ hypothesized to be the set of filler items $F$ by model $m$. $|U_{u^{F_m}}|$ is the number of items in the hypothesized filler partition of profile $P_u$ the mean rating for each of items.

    \item \textbf{Filler Average Correlation (FAC)} measures the correlation between the rating of an item in the hypothesized filler partition of a user profile and the item's average rating:
    \begin{equation}
        \text{FAC}_u=\frac{\sum_{i\in H_u}(r_{u,i}-\bar r_i)}{\sqrt{\sum_{j\in H_u}(r_{u,j}-\bar r_j)^2}}.
    \end{equation}

% \textbf{Filler Average Correlation(FAC)}:The FAC algorithm calculates the correlation between the ratings of the fillers in the profile and the average rating for each item.
% $$
% FAC_u=\frac{\sum_{i\in I_u}(r_{ui}-\bar r_i)}{\sqrt{\sum_{i\in I_u}(r_{ui}-\bar r_i)^2}}
% $$
% where $I_u$ is the set of items rated by user $u$.

    \item \textbf{Mean Variance (MeanVar)} calculates the average variance between the items in the hypothesized filler partition and their average ratings:
    \begin{equation}
    \text{MeanVar}_u=\frac{\sum_{i\in F_u}(r_{u,i}-\bar r_u)^2}{|F_u|},
    \end{equation}
    where $F_u$ contains all the items that user $u$ did not give the maximal rating score $r_{\text{max}}$. For example, $r_{\text{max}}$ is 5 in a five-scale rating system. $\bar{r}_u$ indicates the average rating of all ratings given by $u$.

% \textbf{Mean Variance (MeanVar)}:MeanVar is an iterative process that involves considering each lowly-rated item (for nuke attack) or highly-rated item (for push attack) as a potential target. It calculates the average variance between the filler items and the overall average.
% $$
% MeanVar_u=\frac{\sum_{i\in(R_u-R_{ut})}(r_{ui}-\bar r_u)^2}{|R_u-R_{ut}|}
% $$
% where $R_u$ is the rating vector of user $u$ of all items in rating matrix $R$, and $R_{ut}$ is the collection of item $i$ which $r_{ui}=r_{max}$(or $r_{ui}=r_{min}$ for nuke attack).

    \item \textbf{Filler Mean Target Difference (FMTD)} quantifies the discrepancy between the maximal rating score $r_{\text{max}}$ and rating scores provided by user $u$ that are not maximal: 
    \begin{equation}
    \text{FMTD}_u=\left|\frac{\sum_{i\in M_{u}}r_{\text{max}}}{|M_{u}|}-\frac{\sum_{k\in F_u}r_{u,k}}{|F_u|}\right|,
    \end{equation}
    where $M_{u}$ indicates the items that $u$ gave the maximal rating score $r_{\text{max}}$.

% \textbf{Filler Mean Target Difference(FMTD)}: TMTD serves as a partitioning feature specifically designed for the detection of segment attacks. FMTD focuses on quantifying the discrepancy in ratings provided by user $u$ between items receiving the highest or lowest rating and all other items rated by user $u$ excluding those associated with the extremities. This metric enables the identification of segment attacks with heightened effectiveness. The utilization of FMTD aims to enhance the precision and efficacy of attack detection mechanisms in rating-based systems.
% $$
% FMTD_u=|\frac{\sum_{i\in R_{ut}}r_{ui}}{R_{ut}}-\frac{\sum_{k\in (R_u-R_{ut})}r_{uk}}{|R_u-R_{ut}|}|
% $$

    \item \textbf{Filler Size with Total Items (FSTI)} is the percentage of a user' profile size over the number of items in the RS:
    \begin{equation}
    \text{FSTI}_u=\frac{|N_u|}{|I|},
    \end{equation}
    where $I$ is the item set in RS.

% \textbf{Filler Size with Total Items(FSTI)}: FSTI is the ratio between the profile size of user $u$ and the item-set size in the system dataset.
% $$
% FSTI_u=\frac{\sum_{i\in V}O(r_{ui})}{|V|}
% $$
% where $V$ is the item-set in system dataset, and $O(r_{ui})$ is 1 if user $u$ rated item $i$, 0 otherwise.

    \item \textbf{Filler Size with Popular Items in Itself (FSPII)} is the percentage of most popular items that a user has rated over the profile size:
    \begin{equation}
    \text{FSPII}_u=\frac{\sum_{i\in I_p}\mathbb{I}_1(u,i)}{|N_u|}
    \end{equation}
    where $V_p$ is the most popular items in RS and we define it as the top 5\% most popular items (with many interactions) in RS.
    $\mathbb{I}_1(u,i)$ is 1 if user $u$ has rated item $i$; otherwise 0.

% \textbf{Filler Size with Popular Items in Itself(FSPII)}: the ratio between the number of ratings for popular items of user $u$ and the profile size of user $u$.
% $$
% FSPII_u=\frac{\sum_{i\in V_p}O(r_{ui})}{\sum_{i\in V}O(r_{ui})}
% $$
% where $V_p$ is the popular items in recommender system.

    \item \textbf{Filler Size with Maximum Rating in Itself (FSMAXRI)} indicates the percentage of the times that a user $u$ gave the maximal rating score over $u$'s profile size:
    \begin{equation}
    \text{FSMAXRI}_u=\frac{\sum_{i\in N_u} \mathbb{I}_2(r_{u,i}, r_{\text{max}})}{|N_u|},
    \end{equation} 
    where the indicator $\mathbb{I}_2(r_{u,i}, r_{\text{max}})$ is 1 if $r_{u,i}$ equals $r_{\text{max}}$; otherwise 0.

% \textbf{Filler Size with Maximum Rating in Itself (FSMAXRI)}: FSMAXRI is the number of max ratings score rated by user $u$ and the profile size of user $u$.
% $$
% FSMAXRI_u=\frac{\sum_{i\in V}O(r_{ui}=r_{max})}{\sum_{i\in V}O(r_{ui})}
% $$

    \item \textbf{Filler Size with Average Rating in Itself (FSARI)} indicates the percentage of the times that a user $u$ gave the average rating score over $u$'s profile size:
    \begin{equation}
    \text{FSARI}_u=\frac{\sum_{i\in I}\mathbb{I}_3(r_{u,i}, r_{\text{avg}})}{|N_u|}
    \end{equation} 
    where $r_{\text{avg}}$ is the global average score in RS. The indicator $\mathbb{I}_3(r_{u,i}, r_{\text{avg}})$ is 1 if the floor or the ceiling of $r_{\text{avg}}$ equals $r_{u,i}$; otherwise 0.
 
% \textbf{Filler Size with Average Rating in Itself (FSARI)}: FSARI is the number of average ratings score rated by user $u$ and the profile size of user $u$.
% $$
% FSARI_u=\frac{\sum_{i\in V}O(r_{ui}=r_{avg})}{\sum_{i\in V}O(r_{ui})}
% $$
% where $r_{avg}$ is the average score in the whole system.

\end{enumerate}

Note that, although we illustrate the definitions of the selected features from user side, they can be used as both user features and item features.
Based on the selected features, for each user/item, we construct a normalized 10-dimensional feature vector $\mathbf{x}$.

\subsubsection{Generate Toxic Fake User Features}

Given features of real users and items, the next step is to generate toxic fake user features that can guide the edge generation to fill the fake user profile with user-item interaction data.
To this end, \ours adopts the idea of reconstruction: train a graph encoder by predicting the features of some masked real users and items, and then use the graph encoder to predict the features of the fake user. 

Specifically, \ours first maps features of real users and items into high dimensional representation spaces via a two-layer feedforward neural network:
\begin{equation}
\label{eq:mapping}
\mathbf{p}=\mathbf{W}_2\cdot\text{LeakyRELU}(\mathbf{W}_1 \mathbf{x}),
\end{equation}
where $\mathbf{x}$ is the feature vector of a user or an item, $\mathbf{W}_1$ and $\mathbf{W}_2$ are trainable parameters. 

Then, \ours uses a multi-relation graph convolution layer to aggregate information of neighboring nodes and update representations for each user in the user-item bipartite graph:
\begin{equation}
    \label{eq:agg}
    \begin{aligned}
    \mathbf{q}_u &= \underset{r=1}{\overset {2}{\sum}}\,\, \underset{v\in \mathcal{N}_r (u)}{\sum}\frac {\mathbf{W}_r \mathbf{p}_v}{\sqrt {|\mathcal{N}_r(u)|\cdot|\mathcal{N}_r(v)|}}\\
    \mathbf{h}_u &= \text{LeakyRELU}\big(\mathbf{W}_1\cdot\text{LeakyRELU}(\mathbf{q}_u)\big)
    \end{aligned}
\end{equation}
where $\mathcal{N}_r(u)$ indicates the neighboring node of $u$ w.r.t. to edge type $r$ and there are two types of edges (user$\rightarrow$item edges and item$\rightarrow$user edges).
\ours uses a similar aggregation process for updating item representations.

Next, \ours randomly masks some real user and item nodes and reconstructs the masked features.
\ours uses a two-layer MLP for the feature reconstruction:
\begin{equation}
\label{eq:predict}
    \mathbf{\hat{x}}=\mathbf{W}_{2}^{(rec)} \cdot \sigma\big(\mathbf{W}_{1}^{(rec)} \mathbf{h}\big),
\end{equation}
where $\mathbf{W}_{1}^{(rec)}$ and $\mathbf{W}_{2}^{(rec)}$ are trainable parameters.
Suppose that the masked user set is $U_m$ and the masked item node set is $V_m$, the following reconstruction loss is used for feature reconstruction:
\begin{equation}
    \mathcal{L}_{\text{recon}}=\frac{1}{|U_m|}\underset{u\in U_m}{\sum}||\mathbf{x}_u-\mathbf{\hat x}_u||^2 + \frac{1}{|V_m|}\underset{v\in V_m}{\sum}||\mathbf{x}_v-\mathbf{\hat x}_v||^2,
\end{equation}
where $\mathbf{x}_u$ is the features of user $u \in U_m$ and $\mathbf{\hat x}_u$ is the reconstructed features of $u$. 
Similar notations $\mathbf{x}_v$ and $\mathbf{\hat x}_v$ are used for reconstructing item features.

Through reconstruction, the graph encoder is empowered by the capability to encode topological and feature information of the bipartite graph containing real users and items, and it can be used to generate features for the fake user.
We initialize the feature vector $\mathbf{x}_{z^{\prime}}$ of the fake user $z^{\prime}$ as zero vector and use \ours to predict $\hat{\mathbf{x}}_{z^{\prime}}$ (Equation~\ref{eq:predict}) as fake user features.
However, the generated features for the fake user do not convey toxicity and cannot guide the edge generation to fulfill the attack goal.
Therefore, we further adopt the idea of influence functions~\cite{influence}, a classic technique from robust statistics~\cite{cook1980characterizations} that has shown promising results in determining the importance of a training sample in RS~\cite{TrialAttack}, to endow the generated features of the fake user with destructive power.

To be specific, influence functions show how the model parameters change as we upweight a training sample by an infinitesimal amount.
For a training sample $z$, if it is upweighted by a small value $\epsilon$, the changed parameter $\hat{\theta}_{\epsilon,z}$ can be defined as:
\begin{equation}
\hat{\theta}_{\epsilon,z}\overset{\text{def}}{=}\arg \min_{\theta}\frac{1}{n}\sum_{i=1}^{n}\mathcal{L}_{\text{RS}}(z_i,\theta) + \epsilon\mathcal{L}_{\text{RS}}(z,\theta),
\end{equation}
where $\mathcal{L}(\cdot)_{\text{RS}}$ indicates the training loss of RS and $n$ is the number of samples. 
Then, the influence of upwerighting $z$ on the parameter is given by:
\begin{equation}
\label{eq:inf}
\mathcal{I}_{\text{up},\text{params}}(z)\overset{\text{def}}{=}\frac{d\hat{\theta}_{\epsilon,z}}{d\epsilon}\vert_{\epsilon=0}=-H^{-1}_{\hat{\theta}}\nabla_{\theta}\mathcal{L}_{\text{RS}}(z,\hat{\theta}),
\end{equation}
where $H^{-1}_{\hat{\theta}}=\frac{1}{n}\sum_{i=1}^{n}\nabla_{\theta}\mathcal{L}_{\text{RS}}(z,\hat{\theta})$ is the Hessian matrix of $\mathcal{L}_{\text{RS}}$.
Based on Eq.~\ref{eq:inf}, Koh and Liang~\cite{influence} derivate that the influence of upweighting $z$ on a test sample $z_{\text{test}}$ has a closed-form expression:
\begin{equation}
    \label{eq:influence}
    \mathcal{I}_{\text{up},\text{loss}}(z, z_{test})=-\nabla_{\theta}\mathcal{L}(z_{test}, \hat \theta)^TH^{-1}_{\hat \theta}\nabla_{\theta}\mathcal{L}_{\text{RS}}(z,\hat \theta).
\end{equation}

\hide{

    Considering that the fake user $z^{\prime}$ does not exist in the training data, Wu et al.~\cite{TrialAttack} deduce the formulation of the influence of $z^{\prime}$:
    \begin{equation}
    \label{eq:inf}
    \text{Inf}(z^{\prime})=-\frac1n\nabla_{\theta}\mathcal{L}_{adv}(X,\hat{\theta})^TH_{\hat{\theta}}^{-1}\nabla_z\nabla_{\theta}\mathcal{L}_{\text{RS}}(z_{min},\hat{\theta})(z^{\prime}-z_{min}),
    \end{equation}
    where $H^{-1}_{\hat{\theta}}=\frac{1}{n}\sum_{i=1}^{n}\nabla^{2}_{\theta}\mathcal{L}_{\text{RS}}(z_i,\hat{\theta})$ is the Hessian matrix of $\mathcal{L}_{\text{RS}}$ and $\mathcal{L}_{adv}(X,\hat{\theta})$ indicates the attacking goal which we will shown in Equation~\ref{eq:att_loss}. 
    $z_{min}=\Pi(\mu+e_t)$ where $\Pi{(z)}$ project each $z_i$ to reasonable discrete rating.
    $\mu\in\mathbb{R}^m$ where $m$ is the number of items. 
    Each element $\mu_i$ in $\mu$ represents the rating score on item $i$.
    $r_{max}$ is the maximum rating in RS (e.g., 5 in a five-scale rating system).
    $e\in\mathbb{R}^m$ whose $t$-th dimension is $r_{\max}$ and others are all zero. 

    We follow their method and train a three-layer feedforward neural network as the influence predictor to predict the influence of a user. 
    When training the influence predictor, in each epoch, we randomly sample some real users, and minimize the gap (mean squared error) between their true influence scores (Equation~\ref{eq:inf}) and the predicted influence scores.

    % \hui{At this stage, $z'$ only has generated features. How to use Eq.~\ref{eq:inf}?}

}

We use Equation~\ref{eq:influence} to pre-compute the influence scores of all real users.
We then use a three-layer feedforward neural network as an influence predictor $\text{IP}(\cdot)$.
Given the feature vector $x_z$ of a real user $z$, the influence predictor is trained to predict the influence score of $z$ by minimizing the gap (mean squared error) between the  true influence score and the predicted influence score.

In summary, the optimization objective for generating toxic fake user features can be formulated as:
\begin{equation}
\label{eq:feat_loss}
\mathcal{L}_{\text{feat}} = \mathcal{L}_{\text{recon}} - \text{IP}(x_{z'}).
\end{equation}
Minimizing the loss function in Equation~\ref{eq:feat_loss} trains \ours to reconstruct node features more accurately and maximizing the influence of the fake user $z'$ at the same time.

\subsection{Edge Generation}

An essential step in contemporary shilling attack approaches is filling the fake user profile with interaction history.
This step is equivalent to connecting the fake user node with items in the user-item bipartite graph which ensures the fake user profile can affect the recommendation of the RS on the target item.

Specifically, we use the generated features of the fake user profile to guide the generation of edges.
We first project the predicted fake user features (Equation~\ref{eq:mapping}) and the features of candidate items (i.e., the ten selected RS features) through a single layer feedforward neural network in order to project them to the same space:
\begin{equation}
\mathbf{q}_{z^{\prime}}=\mathbf{W}_{\text{edge}}\hat{\mathbf{x}}_{z^{\prime}},\ \mathbf{q}_{j}=\mathbf{W}_{\text{edge}}\mathbf{x}_{j}
\end{equation}
where $\mathbf{W}_{\text{edge}}$ is trainable parameters and $j$ is a candidate item.
We choose the $2$-hop item neighbors of the target item as the candidate items.
Candidate items and the target item were interacted by same real users in the past.
According to the idea of co-visitation attack~\cite{covisit}, these candidate items can affect whether the target item can be recommended after shilling attack.
To avoid being easily detected, we additionally add $s$ sampled popular items into the candidate set.

Then, we calculate the probability of connecting the fake user to each candidate item by measuring the the cosine similarity between $\mathbf{q}_{z^{\prime}}$ and $\mathbf{q}_{j}$.
The resulting probability distribution $\mathbf{o}\in \mathbb{R}^{b_{\text{item}}}$ contains probabilities of all the candidate items.
Next, our target is to choose top-$b_{\text{item}}$ candidate items with the highest probabilities. 
To address the discretization issue of the network, we employ the Gumbel-Top-K technique. 
It is an extension of the Gumbel-Max trick for sampling from a categorical distribution. 
The Gumbel-Max trick is a method that adds independent and identically distributed (i.i.d.) Gumbel noise to the log-probabilities of each category and selects the category with the highest sum of log-probability and Gumbel noise~\cite{gumbel, efraimidis2006weighted}. 
The Gumbel-Top-k trick extends this method to sample $k$ elements without replacement.  
For $\varepsilon \sim \text{Uniform}(0,1)$, Gumbel-Softmax is defined as:
\begin{equation}
    \text{Gumble-Softmax}(\mathbf{o})_i = \frac{\exp(\frac{(\log(o_i) + g_i)}{\tau})}{\sum_{j=1}^c \exp(\frac{(\log(o_j) + g_j)}{\tau})},
\end{equation}
where $m$ is the size of the candidate item set.
where the parameter $\tau>0$ represents the annealing factor that determines how close the output result is to the one-hot form. 
A smaller $\tau$ value leads to a more one-hot-like output, but may cause a more severe gradient vanishing problem.
$o_i$ is the $i$-th dimension in $\mathbf{o}$. 
The Gumbel distribution $g_i=-\log(-\log\varepsilon_i)$ and it brings exploration to the edge selection process.
And we can further use $\alpha$ to control the strength of exploration:
\begin{equation}
    \text{Gumble-Softmax}(\mathbf{o}, \epsilon)_i = \frac{\exp(\frac{(\log(o_i) + \alpha\cdot g_i)}{\tau})}{\sum_{j=1}^n \exp(\frac{(\log(o_j) + \alpha\cdot g_j)}{\tau})}.
\end{equation}
The Gumbel-Top-K function for edge generation can be formulated as follow:
\begin{equation}
   G(\mathbf{o}) = \overset{b_{\text{item}}}{\underset{i=1}{\sum}}\text{Gumble-Softmax}(\mathbf{o} \odot \text{mask}_i,\alpha)_i,
\end{equation}
where $\text{mask}_i$ filters out the selected edges so that they are not chosen again in subsequent iterations. 
Note that the resulting vector is sharp but not strictly discrete, which facilitates the training process~\cite{single}. 
In the test phase, we enforce a hard threshold $e$ on the vector to choose edges that connect to the fake user.

\subsection{Optimization}

We inject the generated fake user node into the user-item bipartite graph to launch shilling attacks.
We design the following attack loss to endow the generated fake user with strong destructive power:
\begin{equation}
    \label{eq:att_loss}
    \mathcal{L}_{adv}(X, \hat\theta) =- \underset{u\in \mathcal{U}}{\sum}log(\frac{exp(r_{u,t})}{\sum_{j\in\mathcal{I}}exp(r_{u,j})}),
\end{equation}
where $t$ indicates the target item.
Equation~\ref{eq:att_loss} shows the attack goal of promotion shilling attack: hoax RS and mislead RS to rank the target item higher than other items when making recommendation.

In summary, the complete objective of \ours is:
\begin{equation}
    \mathcal{L}=\mathcal{L}_{\text{adv}} + \mathcal{L}_{\text{feat}}=\mathcal{L}_{\text{adv}} + \mathcal{L}_{\text{recon}} - \text{IP}(x_{z'}).
\end{equation}
And \ours can be optimized using gradient descent based methods like Adam~\cite{KingmaB14}. 
When the optimization of \ours finishes, we can construct a fake user in the victim RS and fill the fake user profile with some user-item interactions guided by the generated edges from \ours.
Then, the victim RS will be affected by the injected fake user and the attack goal can be achieved.

%!TEX root = main.tex

\section{Experiment}
\label{sec:exp}

In this section, we present the experimental results and analysis.

\subsection{Experimental Settings}

\begin{table}[t]
    \caption{Statistics of datasets}
    \centering
    \begin{tabular}{ccccc}
    \hline
    Dataset &  Users & Items & Interactions & Sparsity\\ \hline
    Automotive   & 2,928 & 1,835 & 20,473 & 99.62\% \\
    T \& HI & 1,208 & 8,491 & 28,396 & 99.72\% \\
    Last.fm & 1,892 & 12,523 & 186,479 & 99.21\% \\ \hline
    \end{tabular}
    \label{tab:data}
\end{table}

\begin{table*}[t]
    \caption{Attack performance (HR@50) of different attack methods against different victim RS. The left side of the slash is the attack performance for single-user injection, and the right side is the attack performance for multi-user injection. Best results of single-user injection and multi-user injection are shown in bold.}
    \label{tab:results}
    \scalebox{0.83}
    {
    \begin{tabular}{c|c|ccccccccccc}
    \hline 
        \multirow{2}{*}{Dataset} & \multirow{2}{*}{Victim RS} & ~ & ~ & ~ & \multicolumn{3}{c}{Shilling Attack Methods} & ~ & ~ \\ \cline{3-13}
        ~ & ~ & Random & Segment & Bandwagon & TrialAttack & AUSH & LegUP & SUI-Attack & ratio 1 & ratio 2 & ratio 3 & ratio 4 \\ \hline
        ~ & ItemCF & 0.010/0.207 & 0.000/0.126 & 0.000/0.122 & 0.042/\textbf{0.295} & 0.012/0.172 & 0.040/0.253 & \textbf{0.194}/0.262 & 0.658 & 4.620 & 0.740 & 0.888 \\ 
        ~ & WMF & 0.000/0.063 & 0.000/0.020 & 0.000/0.024 & 0.000/\textbf{0.081} & 0.004/0.046 & 0.000/0.050 & \textbf{0.007}/0.062 & 0.086 & 1.750 & 0.113 & 0.765 \\ 
        ~ & NGCF & 0.001/0.093 & 0.001/0.090 & 0.001/\textbf{0.102} & 0.000/0.071 & 0.002/0.090 & 0.000/0.101 & \textbf{0.076}/0.092 & 0.745 & 38.00 & 0.826 & 0.902 \\ 
        T \& HI & VAE & 0.203/0.762 & 0.174/0.826 & 0.000/0.811 & 0.103/\textbf{0.992} & 0.241/0.962 & 0.227/0.931 & \textbf{0.636}/0.937 & 0.641 & 2.639 & 0.679 & 0.945\\ 
        ~ & ItemAE & 0.014/0.234 & 0.001/0.143 & 0.000/0.174 & 0.004/\textbf{0.281} & 0.082/0.145 & 0.002/0.268 & \textbf{0.168}/0.192 & 0.598 & 2.049 & 0.875 & 0.683\\ 
        ~ & LightGCN & 0.000/0.027 & 0.000/0.048 & 0.000/\textbf{0.151} & 0.000/0.039 & 0.000/0.053 & 0.000/0.037 & \textbf{0.023}/0.034 & 0.433 & $+\infty$ & 0.676 & 0.225 \\ 
        ~ & NCF & 0.301/\textbf{0.883} & 0.193/0.717 & 0.163/0.783 & 0.265/0.690 & 0.317/0.804 & 0.187/0.824 & \textbf{0.481}/0.862 & 0.545 & 1.517 & 0.545 & 0.976\\  \hline
        ~ & ItemCF & 0.000/0.201 & 0.002/0.112 & 0.000/0.109 & 0.003/0.211 & 0.003/0.191 & 0.012/\textbf{0.227} & \textbf{0.117}/0.208 & 0.555 & 9.750 & 0.563 & 0.916\\  
        ~ & WMF & 0.001/0.104 & 0.010/0.090 & 0.010/0.163 & 0.095/\textbf{0.201} & 0.020/0.182 & 0.012/0.155 & \textbf{0.118}/0.192 & 0.587 & 1.242 & 0.615 & 0.955\\ 
        ~ & NGCF & 0.084/0.382 & 0.067/0.287 & 0.051/0.248 & \textbf{0.145}/\textbf{0.451} & 0.102/0.414 & 0.094/0.414 & 0.092/0.447 & 0.204 & 0.634 & 0.206 & 0.991\\ 
        Last.FM & VAE & 0.132/0.441 & 0.117/0.372 & 0.163/0.215 & 0.128/0.768 & 0.192/0.537 & 0.118/\textbf{0.854} & \textbf{0.494}/0.683 & 0.578 & 2.573 & 0.578 &0.800\\ 
        ~ & ItemAE & 0.029/\textbf{0.174} & 0.010/0.027 & 0.000/0.084 & 0.004/0.166 & 0.002/0.157 & 0.000/0.147 & \textbf{0.087}/0.162 & 0.524 & 3.000 & 0.537 & 0.931\\ 
        ~ & LightGCN & 0.075/0.182 & 0.010/0.167 & 0.020/0.138 & 0.076/0.314 & 0.091/0.382 & 0.083/\textbf{0.403} & \textbf{0.266}/0.376 & 0.660 & 2.923 & 0.707 & 0.933\\ 
        ~ & NCF & 0.275/0.541 & 0.208/0.462 & 0.164/0.491 & 0.197/0.827 & 0.262/\textbf{0.862} & 0.232/0.817 & \textbf{0.477}/0.835 & 0.553 & 1.735 & 0.571 & 0.969\\  \hline
        ~ & ItemCF & 0.042/0.201 & 0.018/0.184 & 0.000/0.154 & 0.066/\textbf{0.324} & 0.067/0.297 & 0.087/0.313 & \textbf{0.204}/0.307 & 0.630 & 2.345 & 0.664 & 0.948\\ 
        ~ & WMF & 0.019/0.286 & 0.027/0.439 & 0.000/0.337 & 0.082/0.294 & 0.044/0.438 & 0.018/0.398 & \textbf{0.213}/\textbf{0.441} & 0.483 & 2.598 & 0.483 & 1.005\\ 
        ~ & NGCF & 0.010/0.124 & 0.000/0.145 & 0.001/0.119 & 0.000/0.096 & 0.051/\textbf{0.148} & \textbf{0.104}/0.140 & 0.087/0.123 & 0.588 & 0.836 & 0.707 & 0.831\\ 
        Automotive & VAE & 0.030/0.073 & 0.010/0.103 & 0.000/0.096 & 0.010/0.084 & 0.010/\textbf{0.172} & 0.082/0.117 & \textbf{0.091}/0.125 & 0.529 & 1.110 & 0.728 & 0.727\\ 
        ~ & ItemAE & 0.020/0.320 & 0.010/0.176 & 0.002/0.208 & 0.020/0.321 & 0.008/0.311 & 0.091/0.310 & \textbf{0.255}/\textbf{0.322} & 0.792 & 2.802 & 0.792 & 1.003\\ 
        ~ & LightGCN & 0.001/0.141 & 0.002/0.136 & 0.000/0.137 & 0.025/0.184 & 0.018/0.152 & 0.047/0.188 & \textbf{0.162}/\textbf{0.191} & 0.849 & 3.447 & 0.848 & 1.016\\ 
        ~ & NCF & 0.082/0.503 & 0.070/0.515 & 0.002/0.544 & 0.112/0.808 & 0.104/0.762 & 0.142/0.774 & \textbf{0.376}/\textbf{0.811} & 0.464 & 2.648 & 0.464 & 1.004\\ \hline
    \end{tabular}
    }
\end{table*}

\subsubsection{Dataset} 
To demonstrate the effectiveness of our method, we conducted experiments on three public datasets Automotive\footnote{\url{https://github.com/XMUDM/ShillingAttack}\label{auto}}, Tools \& Home Improvement (T \& HI)\textsuperscript{\ref{auto}} and Last.fm\footnote{\url{https://grouplens.org/datasets/hetrec-2011}} that are widely used in previous studies of shilling attacks and RS~\cite{aush,legup,CantadorBK11}. Table~\ref{tab:data} provides the statistics of the data.
We randomly choose 5 items from each dataset as the target items.

\subsubsection{Baselines} 
We compare \ours with traditional shilling attack methods Random Attack~\cite{BurkeMBW05}, Bandwagon Attack~\cite{Burke2005Limited} and Segment Attack~\cite{BurkeMBW05}, and state-of-the-art deep learning based methods, including TrialAttack~\cite{TrialAttack}, AUSH~\cite{aush} and LegUP~\cite{legup}. 
We use the implementation\footnote{\url{https://github.com/ustcml/TrialAttack}} provided by the original authors for TrialAttack.
For other methods, we use the implementations\textsuperscript{\ref{auto}} provided by Lin et al.~\cite{legup}.
We follow the recommended settings of each method.

\subsubsection{Victim RS} 
We conduct shilling attacks against prevalent RS including traditional recommendation models (ItemCF~\cite{SarwarKKR01} and WMF~\citep{HuKV08}) and deep learning based recommendation models (NGCF~\citep{Wang0WFC19}, VAE~\cite{LiangKHJ18}, ItemAE~\cite{SedhainMSX15}, LightGCN~\cite{0001DWLZ020} and NCF~\cite{HeLZNHC17}). We refer to their original papers for parameter settings.

\subsubsection{Evaluation Metric} 
We adopt Hit Ratio (HR@$k$), a metric that is widely used to evaluate the performance of shilling attack~\cite{TrialAttack}. 
It indicates the fraction of users for whom the top-$k$ recommendation list after the attack contains the target item. 
We set $k$ to 50 in our experiments.

\subsubsection{Parameter Settings} 

In feature generation, we randomly mask 10\% of nodes for recovery, the candidate items are 2-hop neighboring items of the target item and the 5\% items (i.e., $s$) sampled from the top 10\% popular items in the RS. 
In edge generation, we set $e=0.85$ to choose the edges that connect to the fake user.
In addition to testing the performance when injecting only one fake user, we also analyze the results when multiple fake users are injected into the RS so that \ours can be compared to contemporary shilling attack methods, and we set the number of fake users ($b_{user}$) to 50 in multi-user injection.
Fake user(s) in both single-user injection and multi-user injection connect to 50 ($b_{item}$) items at most.
We set the training epoch to 64 and batch size to 32. 
The learning rate is set to 5e-4 and we adopt cosine decay learning rate scheduler. 
We use gradient normalization and the max norm is set to 1.0. 
% For influence function, \red{we set $r_d$ to 2 and $r$ to 1}. 
For the graph encoder, we apply a dropout layer to the input of a GCN layer with a dropout rate of 0.5. 
The hidden size of the graph encoder is set to 250. 
We select LeakyReLU as the non-linear activate function with the negative slope being 0.1. 
% For edge generation, we set the projection size $d_r$ to be 64. 

\subsection{Performance of Shilling Attack}

Table~\ref{tab:results} presents the results of our method and the baseline models, and the best results are shown in bold.
The value on the left side of the slash indicates HR@50 of single-user injection using each method. 
For single-user injection, baselines are modified to inject only one fake user.
We also list the results of multi-user injection in the right side of the slash for a comparison.
We provide four types of ratio in Table~\ref{tab:results} for better illustrating the results:
\begin{enumerate}
    \item \textbf{Ratio 1} indicates the percentage of the performance of \ours in single-user injection over the performance of the best baseline in multi-user injection. For example, the ratio 1 for ItemCF on T \& HI is 0.194/0.295=0.658.

    \item \textbf{Ratio 2} represents the percentage of the single-user-injection performance of \ours over the single-user-injection performance of the best baseline. For example, the ratio 2 for ItemCF on T \& HI is 0.194/0.042=4.620.

    \item \textbf{Ratio 3} shows the percentage of single-user-injection performance over multi-user-injection performance of \ours. For example, the ratio 3 for ItemCF on T \& HI is 0.194/0.262=0.740.

    \item \textbf{Ratio 4} is the percentage of the performance of \ours in multi-user injection over the performance of the best baseline in multi-user injection. For example, the ratio 4 for ItemCF on T \& HI is 0.262/0.295=0.888.

\end{enumerate}

From Table~\ref{tab:results}, we have the following findings:
\begin{enumerate}

\item Considering ratio 1, we can also see that \ours which injects only one fake user can generally achieve at least half of the attack performance of the best baseline in multi-user injection, and in some cases ratio 1 can even exceed 0.7.
The observation is encouraging and we find that \textit{even injecting only one fake user can severely mislead the RS to recommend the target item}.
Hence, for some RS where defense mechanisms are deployed, \ours can effectively affect the RS without causing alarm.

\item From ratio 2 shown in Table~\ref{tab:results}, we can observe that, in almost all cases, \ours outperforms baselines by a large margin for single-user injection, suggesting the superiority of \ours in single-user injection. The results also demonstrate that shilling attack methods without tailored designs for single-user injection cannot function well in this challenging setting.

\item When more fake users are injected, we can find that the attack performance of \ours increases (see ratio 3 in Table~\ref{tab:results}) and \ours can achieve comparable or even better performance than contemporary shilling attacks (see ratio 4 in Table~\ref{tab:results}). Therefore, \ours, which is not specifically designed for the traditional shilling attack setting, can work well in multi-user injection, indicating its high flexibility. 

\end{enumerate}

\begin{figure}[t]
    \subfloat[Precision]{%
      \includegraphics[width=0.75\columnwidth]{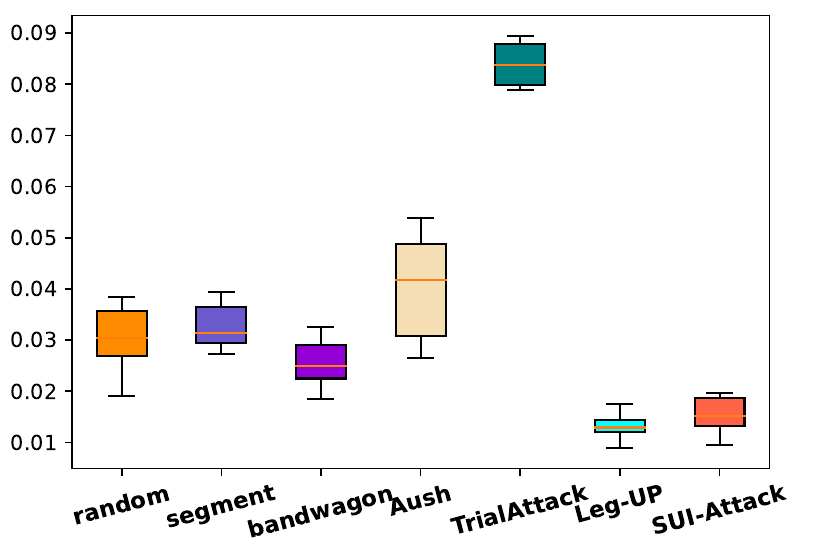}%
    }

    \subfloat[Recall]{%
      \includegraphics[width=0.75\columnwidth]{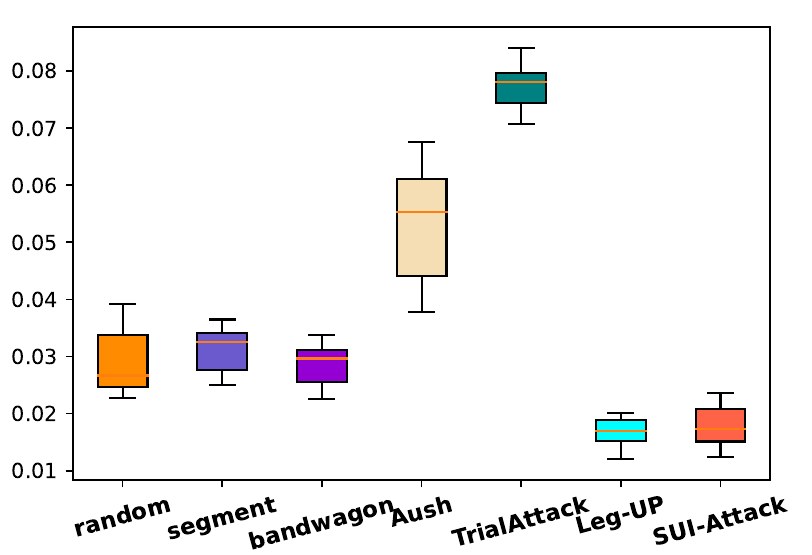}%
    }
\caption{Attack detection of injected profiles on Automotive. Lower value suggests a better attack model.}
\label{fig:detect}
\end{figure}

\begin{figure}[t]
    \subfloat[Last.fm]{%
      \includegraphics[width=0.65\columnwidth]{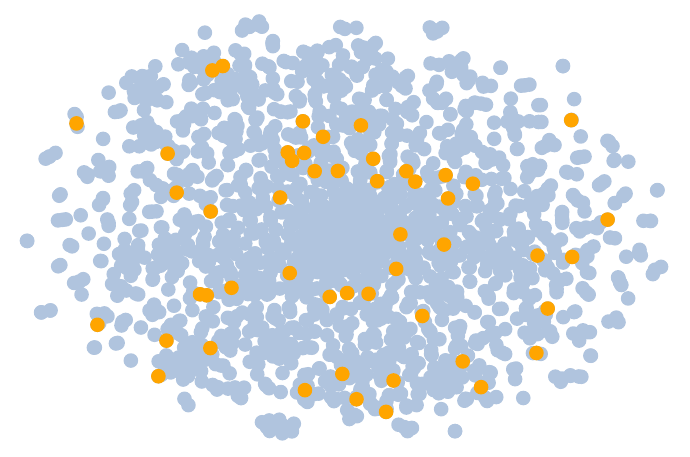}%
    }

    \subfloat[Automotive]{%
      \includegraphics[width=0.65\columnwidth]{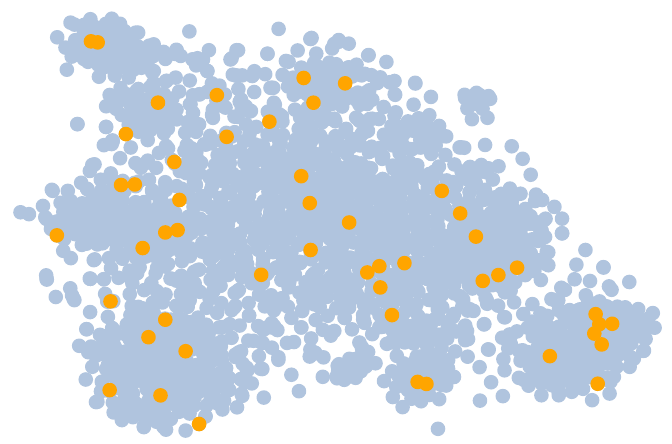}%
    }
\caption{Real and fake user profiles in the latent space. Orange nodes represent injected fake users and other nodes are real users.}
\label{fig:vis}
\end{figure}

\begin{figure}[t]
  \centering
  \includegraphics[width=0.8\columnwidth]{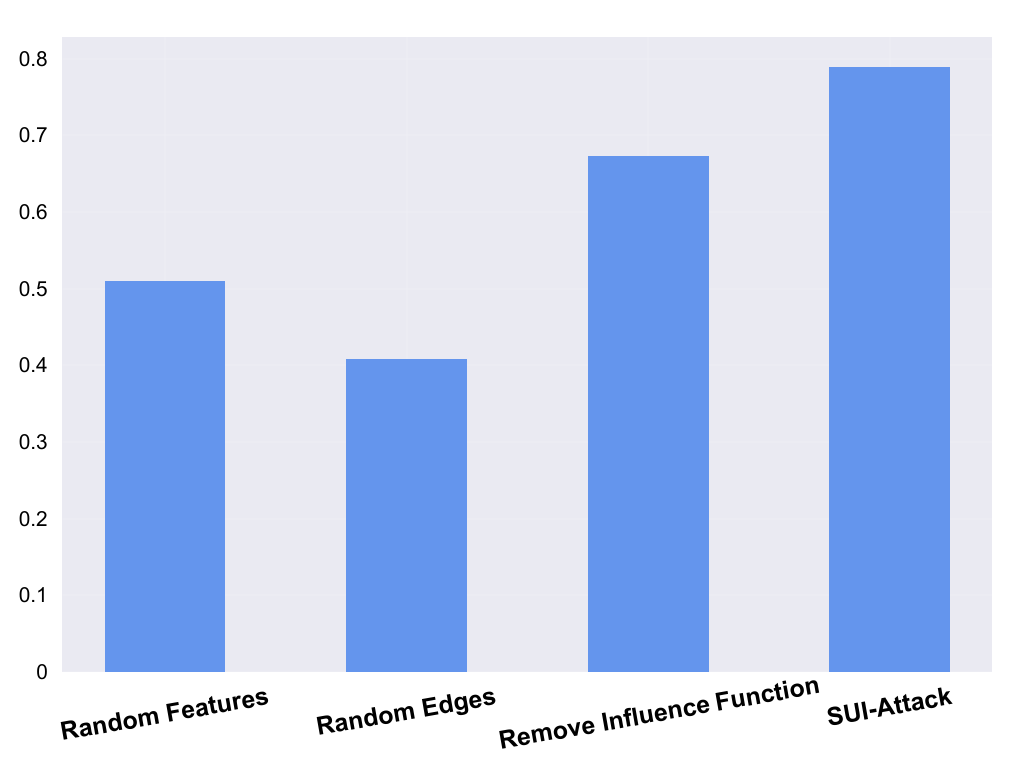}
  \caption{Ablation study.}
  \label{fig:ablation}
\end{figure}

\subsection{Attack Invisibility}

Compared to other shilling attack methods, \ours should be most difficult to detect as it only injects one fake user, the minimum injection for shilling attack, into the victim RS.
Still, we investigate the invisibility of \ours following the study method used by existing shilling attack works~\cite{aush,legup,Zeng0JCL23} in this section.

\subsubsection{Attack Detection}
We use an unsupervised attack detector~\cite{zhang2015catch} to identify the fake user profiles generated by different attack models and report the precision and recall on Automotive in Figure~\ref{fig:detect}. 
Since single-user injection is too difficult to detect, we report the detection results of multi-user injection for \ours.
Lower precision and recall imply that the attack method is more imperceptible. 
The results show that, compared to other attack methods, it is more difficult to detect the fake users generated by \ours.

\subsubsection{Fake User Distribution}
To further study the invisibility of \ours, we visualize the users' representations using the t-SNE projection~\cite{MaatenH08}. 
Note that we visualize the representation space in the multi-user injection as it is meaningless to visualize a single fake user and many real users in single-user injection for checking whether they are different.
Figure~\ref{fig:vis} provides the visualization of users' representations generated by WMF after it is attacked by \ours.
We can observe that fake users are scattered among real users in the representation space and it is hard for detectors to distinguish fake and real users, suggesting that \ours can launch virtually invisible attacks.

\subsection{Ablation Study}

Finally, we discuss the impact of different parts of \ours on the attack performance by conducting ablation experiments. 
Recall that our method mainly consists of two parts: feature generation and edge generation. 
The feature generation process also includes the influence function.
Therefore, our ablation study involves three variants of \ours: 
\begin{enumerate}
    \item \textbf{Replacing feature generation with random feature generation:} Randomly generated fake user features in \ours.

    \item \textbf{Replacing edge generation with random edge generation:} In edge generation, randomly connect the fake user node to other item nodes.

    \item \textbf{Removing the influence function:} It does not use the influence function to guide the generation of toxic fake user features.

\end{enumerate}
Figure~\ref{fig:ablation} shows the performance of the three variants of \ours compared to the performance of the complete \ours on Automotive.
We can clearly see that the three variants show worse attack performance than \ours.
Hence, we can conclude that each part in \ours contributes to its attack performance.

%!TEX root = main.tex

\section{Related Work}
\label{sec:related_work}

In this section, we introduce several directions that are closely related to this work.

\subsection{Recommender Systems (RS)}

The research on RS has a long history~\cite{Aggarwal16}.
Traditional RS typically relies on collaborative filtering (CF) methods,
especially matrix factorization (MF)~\cite{ShiLH14}, where user preferences
and item properties are factorized from the user-item interaction matrix into
two low-dimensional latent matrices. Due to its effectiveness on large-scale
data~\cite{LiCYM17}, MF has been successfully deployed in practice. The
cold-start problem (i.e., data sparsity), where historical data is not
available for new users or items, is one of the most challenging issues in
recommender systems~\cite{Aggarwal16}. To alleviate this problem, additional
context features (e.g., social network~\cite{LiWM14,LiWTM15}, user grouping
data~\cite{DingLHM17,LiLQMTC19},  locations~\cite{LuLMC17}, sequential
data~\cite{LiLMR20}, and review text~\cite{WangNL2019})
are incorporated into MF. 

Recently, the success of deep learning has inspired researchers to deploy deep learning in RS~\cite{WuHWZW23}.
Various prevalent deep learning techniques have been applied in RS.
For instance, RecSeats~\cite{MoinsAB20} adopts Convolutional Neural Network (CNN) in seat recommendation, Zhou et al.~\cite{abs-2202-13556} uses Multilayer Perceptron (MLP) to enhance recommendation, Zhang et al.~\cite{zhang2022deep} deploy Recurrent Neural Network (RNN) to capture the local/global sessions within sequences for CTR prediction, Wang et al.~\cite{WangNL18} leverage Generative Adversarial Network (GAN)~\citep{GoodfellowPMXWOCB14} in cross-domain recommendation, and Li et al.~\cite{0057LX0LJ22} harness Graph Neural Network (GNN) to model social recommendation.
The use of deep learning methods has significantly improved the quality of recommendation~\cite{ZhangYST19}.

Due to the importance of RS for guiding users towards making decisions, RS have attracted unscrupulous parties and there exist various types of attacks against RS in the literature, including
unorganized malicious attacks (i.e., several attackers individually attack RS without an organizer)~\cite{Pang0TZ18}, sybil attacks (i.e., attacker illegally
infers a user's preference)~\cite{CalandrinoKNFS11}, shilling attack, etc. 

\subsection{Shilling Attack against RS}

In the literature, shilling attack is also called as data poisoning attack~\cite{li2016data,ChenL19} or profile injection attack~\cite{BurkeMBW05}.
In experiments, previous works have successfully performed shilling attacks against real-world RS such as YouTube, Google Search, Amazon and Yelp~\cite{XingMDSFL13,covisit}. 
Sony, Amazon and eBay have also reported that they suffered from shilling attacks~\cite{LamR04}.

Pioneering shilling attack methods mainly rely on heuristics and data statistics.
Lam and Riedl~\cite{LamR04}, Burke et al.~\cite{BurkeMBW05,Burke2005Limited} and Mobasher et al.~\cite{d3} 
propose several heuristic based shilling attack approaches to promote an item (e.g., 
Random, Average, Bandwagon and Segment Attacks) or demote an item (e.g., Love/Hate Attacks and Reverse Bandwagon Attacks) for both rating prediction and top-$K$ recommendation. 
Wilson and Seminario~\cite{WilsonS13,SeminarioW14b} propose power user attack and power item attack which leverage most influential users/items to hoax RS.
Fang et al.~\cite{FangYGL18} study shilling attack methods to spoof graph based RS. 
Li et al.~\cite{li2016data} present shilling attack method against factorization based RS. 
Xing et al.~\cite{XingMDSFL13} and Yang et al.~\cite{covisit} conduct experiments on attacking real-world RS (e.g., YouTube and Amazon), and the results show that attacking RS is possible in practice. 

Recently, there is a surge of works on adversarial attack against text and image based
learning systems~\cite{Zhang20,YuanHZL19} and they show that, crafted adversarial examples, which may be imperceptible, can lead to unexpected mistakes of machine learning based systems. 
Based on the idea of adversarial attack, a great number of shilling attack approaches have sprung up.
Optimization based methods~\cite{revisit,li2016data,ZhangTLSYZG21} model shilling attacks as an optimization task and then design optimization strategies to solve it. 
GAN based methods~\cite{aush,legup,TrialAttack,christakopoulou2019adversarial,ZhangCZWL21} use GAN to construct fake user profiles.
Reinforcement learning based methods~\cite{SongLHWLLG20,ZhangLD020,FanDZ0LWT021} query the RS to get feedback on the attack. Then, they use Reinforcement Learning (RL)~\citep{KaelblingLM96} to adjust the injection.
Knowledge distillation based methods~\cite{YueHZM21} and pre-training based methods~\cite{Zeng0JCL23} are designed to reduce the requirement of prior knowledge and improve the practicality of shilling attack.

Although many shilling attack methods exist, they all adopt the same attack paradigm, i.e., multi-user injection.
None of them consider the extremely limited scenario, single-user injection, that studied in this work.

\subsection{Adversarial Attacks in the Extremely Limited Scenarios}
The idea of attacking a machine learning model by altering only one element of the input was first proposed in computer vision domain. 
Su et al~\cite{su2019one} propose one-pixel attack and show that it can achieve high success rate when changing just one pixel to make the image misclassified by image classification algorithms. 
This work initiates the discussion of adversarial learning in extremely limited scenarios~\cite{one1,one2}. 
Recently, Finkelshtein et al.~\cite{finkelshtein2022single} and Tao et al.~\cite{single} extend this idea to adversarial learning in graph representation learning. 
Finkelshtein et al.~\cite{finkelshtein2022single} shows that GNNs can be fooled by only slightly perturbing the features or the neighbor list of a single arbitrary node.
The attack is effective even when the attacker cannot choose which node to perturb, and even when GNNs are trained with robust optimization techniques.
Tao et al.~\cite{single} demonstrate that GNNs can be misled by a single injected node to misclassify the target node (i.e., single-node injection attack). 

%!TEX root = main.tex

\section{Conclusion}
\label{sec:con}

In this paper, we investigate a challenging scenario of shilling attack where only one fake user is injected into RS to launch the attack. 
We reformulate the shilling attack problem as a node generation task over the user-item bipartite graph of RS, which enables us to leverage more information in RS to construct the fake user profile. 
We propose \ours, the first shilling attack method that can be used in single-user injection.
Experiments show that \ours can achieve promising attack results in single-user injection. 
Moreover, in the traditional multi-user injection setting, \ours is shown to be effective and can cause comparable attack results compared to existing shilling attack methods, showing its flexibility in shilling attacks.
In the future, we will explore the underlying mechanism of the successful attack with a single injected node and try to design defense strategies against our \ours.

\begin{acks}
This work was partially supported by National Key R\&D Program of China (No. 2022ZD0118201), National Natural Science Foundation of China (No. 62002303, 42171456), and Natural Science Foundation of Fujian Province of China (No. 2020J05001).
\end{acks}

\clearpage
\bibliographystyle{ACM-Reference-Format}
\balance
\bibliography{reference} 


\begin{thebibliography}{72}

%%% the Web address.

\ifx \showCODEN    \undefined \def \showCODEN     #1{\unskip}     \fi
\ifx \showDOI      \undefined \def \showDOI       #1{#1}\fi
\ifx \showISBNx    \undefined \def \showISBNx     #1{\unskip}     \fi
\ifx \showISBNxiii \undefined \def \showISBNxiii  #1{\unskip}     \fi
\ifx \showISSN     \undefined \def \showISSN      #1{\unskip}     \fi
\ifx \showLCCN     \undefined \def \showLCCN      #1{\unskip}     \fi
\ifx \shownote     \undefined \def \shownote      #1{#1}          \fi
\ifx \showarticletitle \undefined \def \showarticletitle #1{#1}   \fi
\ifx \showURL      \undefined \def \showURL       {\relax}        \fi
% invisible to TeX
\providecommand\bibfield[2]{#2}
\providecommand\bibinfo[2]{#2}
\providecommand\natexlab[1]{#1}
\providecommand\showeprint[2][]{arXiv:#2}

\bibitem[Aggarwal(2016)]%
        {Aggarwal16}
\bibfield{author}{\bibinfo{person}{Charu~C. Aggarwal}.}
  \bibinfo{year}{2016}\natexlab{}.
\newblock \bibinfo{booktitle}{\emph{Recommender Systems - The Textbook}}.
\newblock \bibinfo{publisher}{Springer}.
\newblock


\bibitem[Akhtar and Mian(2018)]%
        {one1}
\bibfield{author}{\bibinfo{person}{Naveed Akhtar} {and}
  \bibinfo{person}{Ajmal~S. Mian}.} \bibinfo{year}{2018}\natexlab{}.
\newblock \showarticletitle{Threat of Adversarial Attacks on Deep Learning in
  Computer Vision: {A} Survey}.
\newblock \bibinfo{journal}{\emph{{IEEE} Access}}  \bibinfo{volume}{6}
  (\bibinfo{year}{2018}), \bibinfo{pages}{14410--14430}.
\newblock


\bibitem[Borchers et~al\mbox{.}(1998)]%
        {BorchersHR98}
\bibfield{author}{\bibinfo{person}{Al Borchers}, \bibinfo{person}{Jonathan~L.
  Herlocker}, {and} \bibinfo{person}{John Riedl}.}
  \bibinfo{year}{1998}\natexlab{}.
\newblock \showarticletitle{Ganging up on Information Overload}.
\newblock \bibinfo{journal}{\emph{Computer}} \bibinfo{volume}{31},
  \bibinfo{number}{4} (\bibinfo{year}{1998}), \bibinfo{pages}{106--108}.
\newblock


\bibitem[Burke et~al\mbox{.}(2005a)]%
        {Burke2005Limited}
\bibfield{author}{\bibinfo{person}{Robin Burke}, \bibinfo{person}{Bamshad
  Mobasher}, {and} \bibinfo{person}{Runa Bhaumik}.}
  \bibinfo{year}{2005}\natexlab{a}.
\newblock \showarticletitle{Limited Knowledge Shilling Attacks in Collaborative
  Filtering Systems}. In \bibinfo{booktitle}{\emph{{ITWP@IJCAI}}}.
\newblock


\bibitem[Burke et~al\mbox{.}(2005b)]%
        {BurkeMBW05}
\bibfield{author}{\bibinfo{person}{Robin~D. Burke}, \bibinfo{person}{Bamshad
  Mobasher}, \bibinfo{person}{Runa Bhaumik}, {and} \bibinfo{person}{Chad
  Williams}.} \bibinfo{year}{2005}\natexlab{b}.
\newblock \showarticletitle{Segment-Based Injection Attacks against
  Collaborative Filtering Recommender Systems}. In
  \bibinfo{booktitle}{\emph{{ICDM}}}. \bibinfo{pages}{577--580}.
\newblock


\bibitem[Calandrino et~al\mbox{.}(2011)]%
        {CalandrinoKNFS11}
\bibfield{author}{\bibinfo{person}{Joseph~A. Calandrino}, \bibinfo{person}{Ann
  Kilzer}, \bibinfo{person}{Arvind Narayanan}, \bibinfo{person}{Edward~W.
  Felten}, {and} \bibinfo{person}{Vitaly Shmatikov}.}
  \bibinfo{year}{2011}\natexlab{}.
\newblock \showarticletitle{``You Might Also Like:'' Privacy Risks of
  Collaborative Filtering}. In \bibinfo{booktitle}{\emph{{IEEE} Symposium on
  Security and Privacy}}. \bibinfo{pages}{231--246}.
\newblock


\bibitem[Cantador et~al\mbox{.}(2011)]%
        {CantadorBK11}
\bibfield{author}{\bibinfo{person}{Iv{\'{a}}n Cantador}, \bibinfo{person}{Peter
  Brusilovsky}, {and} \bibinfo{person}{Tsvi Kuflik}.}
  \bibinfo{year}{2011}\natexlab{}.
\newblock \showarticletitle{Second workshop on information heterogeneity and
  fusion in recommender systems (HetRec2011)}. In
  \bibinfo{booktitle}{\emph{RecSys}}. \bibinfo{pages}{387--388}.
\newblock


\bibitem[Chen and Li(2019)]%
        {ChenL19}
\bibfield{author}{\bibinfo{person}{Huiyuan Chen} {and} \bibinfo{person}{Jing
  Li}.} \bibinfo{year}{2019}\natexlab{}.
\newblock \showarticletitle{Data Poisoning Attacks on Cross-domain
  Recommendation}. In \bibinfo{booktitle}{\emph{{CIKM}}}.
  \bibinfo{pages}{2177--2180}.
\newblock


\bibitem[Christakopoulou and Banerjee(2019)]%
        {christakopoulou2019adversarial}
\bibfield{author}{\bibinfo{person}{Konstantina Christakopoulou} {and}
  \bibinfo{person}{Arindam Banerjee}.} \bibinfo{year}{2019}\natexlab{}.
\newblock \showarticletitle{Adversarial attacks on an oblivious recommender}.
  In \bibinfo{booktitle}{\emph{RecSys}}. \bibinfo{pages}{322--330}.
\newblock


\bibitem[Cook and Weisberg(1980)]%
        {cook1980characterizations}
\bibfield{author}{\bibinfo{person}{R~Dennis Cook} {and}
  \bibinfo{person}{Sanford Weisberg}.} \bibinfo{year}{1980}\natexlab{}.
\newblock \showarticletitle{Characterizations of an empirical influence
  function for detecting influential cases in regression}.
\newblock \bibinfo{journal}{\emph{Technometrics}} \bibinfo{volume}{22},
  \bibinfo{number}{4} (\bibinfo{year}{1980}), \bibinfo{pages}{495--508}.
\newblock


\bibitem[Deldjoo et~al\mbox{.}(2022)]%
        {DeldjooNM21}
\bibfield{author}{\bibinfo{person}{Yashar Deldjoo}, \bibinfo{person}{Tommaso~Di
  Noia}, {and} \bibinfo{person}{Felice~Antonio Merra}.}
  \bibinfo{year}{2022}\natexlab{}.
\newblock \showarticletitle{A Survey on Adversarial Recommender Systems: From
  Attack/Defense Strategies to Generative Adversarial Networks}.
\newblock \bibinfo{journal}{\emph{{ACM} Comput. Surv.}} \bibinfo{volume}{54},
  \bibinfo{number}{2} (\bibinfo{year}{2022}), \bibinfo{pages}{35:1--35:38}.
\newblock


\bibitem[Ding et~al\mbox{.}(2017)]%
        {DingLHM17}
\bibfield{author}{\bibinfo{person}{Danhao Ding}, \bibinfo{person}{Hui Li},
  \bibinfo{person}{Zhipeng Huang}, {and} \bibinfo{person}{Nikos Mamoulis}.}
  \bibinfo{year}{2017}\natexlab{}.
\newblock \showarticletitle{Efficient Fault-Tolerant Group Recommendation Using
  alpha-beta-core}. In \bibinfo{booktitle}{\emph{{CIKM}}}.
  \bibinfo{pages}{2047--2050}.
\newblock


\bibitem[Efraimidis and Spirakis(2006)]%
        {efraimidis2006weighted}
\bibfield{author}{\bibinfo{person}{Pavlos~S. Efraimidis} {and}
  \bibinfo{person}{Paul~G. Spirakis}.} \bibinfo{year}{2006}\natexlab{}.
\newblock \showarticletitle{Weighted random sampling with a reservoir}.
\newblock \bibinfo{journal}{\emph{Inf. Process. Lett.}} \bibinfo{volume}{97},
  \bibinfo{number}{5} (\bibinfo{year}{2006}), \bibinfo{pages}{181--185}.
\newblock


\bibitem[Fan et~al\mbox{.}(2021)]%
        {FanDZ0LWT021}
\bibfield{author}{\bibinfo{person}{Wenqi Fan}, \bibinfo{person}{Tyler Derr},
  \bibinfo{person}{Xiangyu Zhao}, \bibinfo{person}{Yao Ma},
  \bibinfo{person}{Hui Liu}, \bibinfo{person}{Jianping Wang},
  \bibinfo{person}{Jiliang Tang}, {and} \bibinfo{person}{Qing Li}.}
  \bibinfo{year}{2021}\natexlab{}.
\newblock \showarticletitle{Attacking Black-box Recommendations via Copying
  Cross-domain User Profiles}. In \bibinfo{booktitle}{\emph{{ICDE}}}.
  \bibinfo{pages}{1583--1594}.
\newblock


\bibitem[Fang et~al\mbox{.}(2018)]%
        {FangYGL18}
\bibfield{author}{\bibinfo{person}{Minghong Fang}, \bibinfo{person}{Guolei
  Yang}, \bibinfo{person}{Neil~Zhenqiang Gong}, {and} \bibinfo{person}{Jia
  Liu}.} \bibinfo{year}{2018}\natexlab{}.
\newblock \showarticletitle{Poisoning Attacks to Graph-Based Recommender
  Systems}. In \bibinfo{booktitle}{\emph{{ACSAC}}}. \bibinfo{pages}{381--392}.
\newblock


\bibitem[Finkelshtein et~al\mbox{.}(2022)]%
        {finkelshtein2022single}
\bibfield{author}{\bibinfo{person}{Ben Finkelshtein}, \bibinfo{person}{Chaim
  Baskin}, \bibinfo{person}{Evgenii Zheltonozhskii}, {and} \bibinfo{person}{Uri
  Alon}.} \bibinfo{year}{2022}\natexlab{}.
\newblock \showarticletitle{Single-node attacks for fooling graph neural
  networks}.
\newblock \bibinfo{journal}{\emph{Neurocomputing}}  \bibinfo{volume}{513}
  (\bibinfo{year}{2022}), \bibinfo{pages}{1--12}.
\newblock


\bibitem[Goodfellow et~al\mbox{.}(2014)]%
        {GoodfellowPMXWOCB14}
\bibfield{author}{\bibinfo{person}{Ian~J. Goodfellow}, \bibinfo{person}{Jean
  Pouget{-}Abadie}, \bibinfo{person}{Mehdi Mirza}, \bibinfo{person}{Bing Xu},
  \bibinfo{person}{David Warde{-}Farley}, \bibinfo{person}{Sherjil Ozair},
  \bibinfo{person}{Aaron~C. Courville}, {and} \bibinfo{person}{Yoshua Bengio}.}
  \bibinfo{year}{2014}\natexlab{}.
\newblock \showarticletitle{Generative Adversarial Nets}. In
  \bibinfo{booktitle}{\emph{{NIPS}}}. \bibinfo{pages}{2672--2680}.
\newblock


\bibitem[Gunes et~al\mbox{.}(2014)]%
        {attack_survey}
\bibfield{author}{\bibinfo{person}{Ihsan Gunes}, \bibinfo{person}{Cihan
  Kaleli}, \bibinfo{person}{Alper Bilge}, {and} \bibinfo{person}{Huseyin
  Polat}.} \bibinfo{year}{2014}\natexlab{}.
\newblock \showarticletitle{Shilling attacks against recommender systems: a
  comprehensive survey}.
\newblock \bibinfo{journal}{\emph{Artif. Intell. Rev.}} \bibinfo{volume}{42},
  \bibinfo{number}{4} (\bibinfo{year}{2014}), \bibinfo{pages}{767--799}.
\newblock


\bibitem[He et~al\mbox{.}(2020)]%
        {0001DWLZ020}
\bibfield{author}{\bibinfo{person}{Xiangnan He}, \bibinfo{person}{Kuan Deng},
  \bibinfo{person}{Xiang Wang}, \bibinfo{person}{Yan Li},
  \bibinfo{person}{Yong{-}Dong Zhang}, {and} \bibinfo{person}{Meng Wang}.}
  \bibinfo{year}{2020}\natexlab{}.
\newblock \showarticletitle{LightGCN: Simplifying and Powering Graph
  Convolution Network for Recommendation}. In
  \bibinfo{booktitle}{\emph{{SIGIR}}}. \bibinfo{pages}{639--648}.
\newblock


\bibitem[He et~al\mbox{.}(2017)]%
        {HeLZNHC17}
\bibfield{author}{\bibinfo{person}{Xiangnan He}, \bibinfo{person}{Lizi Liao},
  \bibinfo{person}{Hanwang Zhang}, \bibinfo{person}{Liqiang Nie},
  \bibinfo{person}{Xia Hu}, {and} \bibinfo{person}{Tat{-}Seng Chua}.}
  \bibinfo{year}{2017}\natexlab{}.
\newblock \showarticletitle{Neural Collaborative Filtering}. In
  \bibinfo{booktitle}{\emph{{WWW}}}. \bibinfo{pages}{173--182}.
\newblock


\bibitem[Hu et~al\mbox{.}(2008)]%
        {HuKV08}
\bibfield{author}{\bibinfo{person}{Yifan Hu}, \bibinfo{person}{Yehuda Koren},
  {and} \bibinfo{person}{Chris Volinsky}.} \bibinfo{year}{2008}\natexlab{}.
\newblock \showarticletitle{Collaborative Filtering for Implicit Feedback
  Datasets}. In \bibinfo{booktitle}{\emph{{ICDM}}}. \bibinfo{pages}{263--272}.
\newblock


\bibitem[Jang et~al\mbox{.}(2017)]%
        {gumbel}
\bibfield{author}{\bibinfo{person}{Eric Jang}, \bibinfo{person}{Shixiang Gu},
  {and} \bibinfo{person}{Ben Poole}.} \bibinfo{year}{2017}\natexlab{}.
\newblock \showarticletitle{Categorical Reparameterization with
  Gumbel-Softmax}. In \bibinfo{booktitle}{\emph{{ICLR} (Poster)}}.
\newblock
\urldef\tempurl%
\url{https://openreview.net/pdf?id=rkE3y85ee}
\showURL{%
\tempurl}


\bibitem[Kaelbling et~al\mbox{.}(1996)]%
        {KaelblingLM96}
\bibfield{author}{\bibinfo{person}{Leslie~Pack Kaelbling},
  \bibinfo{person}{Michael~L. Littman}, {and} \bibinfo{person}{Andrew~W.
  Moore}.} \bibinfo{year}{1996}\natexlab{}.
\newblock \showarticletitle{Reinforcement Learning: {A} Survey}.
\newblock \bibinfo{journal}{\emph{J. Artif. Intell. Res.}}  \bibinfo{volume}{4}
  (\bibinfo{year}{1996}), \bibinfo{pages}{237--285}.
\newblock


\bibitem[Kingma and Ba(2015)]%
        {KingmaB14}
\bibfield{author}{\bibinfo{person}{Diederik~P. Kingma} {and}
  \bibinfo{person}{Jimmy Ba}.} \bibinfo{year}{2015}\natexlab{}.
\newblock \showarticletitle{Adam: {A} Method for Stochastic Optimization}. In
  \bibinfo{booktitle}{\emph{{ICLR} (Poster)}}.
\newblock
\urldef\tempurl%
\url{https://arxiv.org/abs/1412.6980}
\showURL{%
\tempurl}


\bibitem[Koh and Liang(2017)]%
        {influence}
\bibfield{author}{\bibinfo{person}{Pang~Wei Koh} {and} \bibinfo{person}{Percy
  Liang}.} \bibinfo{year}{2017}\natexlab{}.
\newblock \showarticletitle{Understanding Black-box Predictions via Influence
  Functions}. In \bibinfo{booktitle}{\emph{{ICML}}}, Vol.~\bibinfo{volume}{70}.
  \bibinfo{pages}{1885--1894}.
\newblock


\bibitem[K{\"{u}}gler et~al\mbox{.}(2018)]%
        {one2}
\bibfield{author}{\bibinfo{person}{David K{\"{u}}gler},
  \bibinfo{person}{Alexander Distergoft}, \bibinfo{person}{Arjan Kuijper},
  {and} \bibinfo{person}{Anirban Mukhopadhyay}.}
  \bibinfo{year}{2018}\natexlab{}.
\newblock \showarticletitle{Exploring Adversarial Examples - Patterns of
  One-Pixel Attacks}. In \bibinfo{booktitle}{\emph{MLCN/DLF/iMIMIC@MICCAI}},
  Vol.~\bibinfo{volume}{11038}. \bibinfo{pages}{70--78}.
\newblock


\bibitem[Lam and Riedl(2004)]%
        {LamR04}
\bibfield{author}{\bibinfo{person}{Shyong~K. Lam} {and} \bibinfo{person}{John
  Riedl}.} \bibinfo{year}{2004}\natexlab{}.
\newblock \showarticletitle{Shilling recommender systems for fun and profit}.
  In \bibinfo{booktitle}{\emph{{WWW}}}. \bibinfo{pages}{393--402}.
\newblock


\bibitem[Li et~al\mbox{.}(2016)]%
        {li2016data}
\bibfield{author}{\bibinfo{person}{Bo Li}, \bibinfo{person}{Yining Wang},
  \bibinfo{person}{Aarti Singh}, {and} \bibinfo{person}{Yevgeniy Vorobeychik}.}
  \bibinfo{year}{2016}\natexlab{}.
\newblock \showarticletitle{Data Poisoning Attacks on Factorization-Based
  Collaborative Filtering}. In \bibinfo{booktitle}{\emph{{NIPS}}}.
  \bibinfo{pages}{1885--1893}.
\newblock


\bibitem[Li et~al\mbox{.}(2017)]%
        {LiCYM17}
\bibfield{author}{\bibinfo{person}{Hui Li}, \bibinfo{person}{Tsz~Nam Chan},
  \bibinfo{person}{Man~Lung Yiu}, {and} \bibinfo{person}{Nikos Mamoulis}.}
  \bibinfo{year}{2017}\natexlab{}.
\newblock \showarticletitle{{FEXIPRO:} Fast and Exact Inner Product Retrieval
  in Recommender Systems}. In \bibinfo{booktitle}{\emph{{SIGMOD} Conference}}.
  \bibinfo{pages}{835--850}.
\newblock


\bibitem[Li et~al\mbox{.}(2022)]%
        {0057LX0LJ22}
\bibfield{author}{\bibinfo{person}{Hui Li}, \bibinfo{person}{Lianyun Li},
  \bibinfo{person}{Guipeng Xv}, \bibinfo{person}{Chen Lin}, \bibinfo{person}{Ke
  Li}, {and} \bibinfo{person}{Bingchuan Jiang}.}
  \bibinfo{year}{2022}\natexlab{}.
\newblock \showarticletitle{{SPEX:} {A} Generic Framework for Enhancing Neural
  Social Recommendation}.
\newblock \bibinfo{journal}{\emph{{ACM} Trans. Inf. Syst.}}
  \bibinfo{volume}{40}, \bibinfo{number}{2} (\bibinfo{year}{2022}),
  \bibinfo{pages}{37:1--37:33}.
\newblock


\bibitem[Li et~al\mbox{.}(2020)]%
        {LiLMR20}
\bibfield{author}{\bibinfo{person}{Hui Li}, \bibinfo{person}{Ye Liu},
  \bibinfo{person}{Nikos Mamoulis}, {and} \bibinfo{person}{David~S.
  Rosenblum}.} \bibinfo{year}{2020}\natexlab{}.
\newblock \showarticletitle{Translation-Based Sequential Recommendation for
  Complex Users on Sparse Data}.
\newblock \bibinfo{journal}{\emph{{IEEE} Trans. Knowl. Data Eng.}}
  \bibinfo{volume}{32}, \bibinfo{number}{8} (\bibinfo{year}{2020}),
  \bibinfo{pages}{1639--1651}.
\newblock


\bibitem[Li et~al\mbox{.}(2019)]%
        {LiLQMTC19}
\bibfield{author}{\bibinfo{person}{Hui Li}, \bibinfo{person}{Yu Liu},
  \bibinfo{person}{Yuqiu Qian}, \bibinfo{person}{Nikos Mamoulis},
  \bibinfo{person}{Wenting Tu}, {and} \bibinfo{person}{David~W. Cheung}.}
  \bibinfo{year}{2019}\natexlab{}.
\newblock \showarticletitle{{HHMF:} hidden hierarchical matrix factorization
  for recommender systems}.
\newblock \bibinfo{journal}{\emph{Data Min. Knowl. Discov.}}
  \bibinfo{volume}{33}, \bibinfo{number}{6} (\bibinfo{year}{2019}),
  \bibinfo{pages}{1548--1582}.
\newblock


\bibitem[Li et~al\mbox{.}(2014)]%
        {LiWM14}
\bibfield{author}{\bibinfo{person}{Hui Li}, \bibinfo{person}{Dingming Wu},
  {and} \bibinfo{person}{Nikos Mamoulis}.} \bibinfo{year}{2014}\natexlab{}.
\newblock \showarticletitle{A revisit to social network-based recommender
  systems}. In \bibinfo{booktitle}{\emph{{SIGIR}}}.
  \bibinfo{pages}{1239--1242}.
\newblock


\bibitem[Li et~al\mbox{.}(2015)]%
        {LiWTM15}
\bibfield{author}{\bibinfo{person}{Hui Li}, \bibinfo{person}{Dingming Wu},
  \bibinfo{person}{Wenbin Tang}, {and} \bibinfo{person}{Nikos Mamoulis}.}
  \bibinfo{year}{2015}\natexlab{}.
\newblock \showarticletitle{Overlapping Community Regularization for Rating
  Prediction in Social Recommender Systems}. In
  \bibinfo{booktitle}{\emph{RecSys}}. \bibinfo{pages}{27--34}.
\newblock


\bibitem[Liang et~al\mbox{.}(2018)]%
        {LiangKHJ18}
\bibfield{author}{\bibinfo{person}{Dawen Liang}, \bibinfo{person}{Rahul~G.
  Krishnan}, \bibinfo{person}{Matthew~D. Hoffman}, {and} \bibinfo{person}{Tony
  Jebara}.} \bibinfo{year}{2018}\natexlab{}.
\newblock \showarticletitle{Variational Autoencoders for Collaborative
  Filtering}. In \bibinfo{booktitle}{\emph{{WWW}}}. \bibinfo{pages}{689--698}.
\newblock


\bibitem[Lin et~al\mbox{.}(2020)]%
        {aush}
\bibfield{author}{\bibinfo{person}{Chen Lin}, \bibinfo{person}{Si Chen},
  \bibinfo{person}{Hui Li}, \bibinfo{person}{Yanghua Xiao},
  \bibinfo{person}{Lianyun Li}, {and} \bibinfo{person}{Qian Yang}.}
  \bibinfo{year}{2020}\natexlab{}.
\newblock \showarticletitle{Attacking Recommender Systems with Augmented User
  Profiles}. In \bibinfo{booktitle}{\emph{{CIKM}}}. \bibinfo{pages}{855--864}.
\newblock


\bibitem[Lin et~al\mbox{.}(2022)]%
        {legup}
\bibfield{author}{\bibinfo{person}{Chen Lin}, \bibinfo{person}{Si Chen},
  \bibinfo{person}{Meifang Zeng}, \bibinfo{person}{Sheng Zhang},
  \bibinfo{person}{Min Gao}, {and} \bibinfo{person}{Hui Li}.}
  \bibinfo{year}{2022}\natexlab{}.
\newblock \showarticletitle{Shilling Black-box Recommender Systems by Learning
  to Generate Fake User Profiles}.
\newblock \bibinfo{journal}{\emph{arXiv Preprint}} (\bibinfo{year}{2022}).
\newblock
\urldef\tempurl%
\url{https://arxiv.org/pdf/2206.11433.pdf}
\showURL{%
\tempurl}


\bibitem[Lu et~al\mbox{.}(2017)]%
        {LuLMC17}
\bibfield{author}{\bibinfo{person}{Ziyu Lu}, \bibinfo{person}{Hui Li},
  \bibinfo{person}{Nikos Mamoulis}, {and} \bibinfo{person}{David~W. Cheung}.}
  \bibinfo{year}{2017}\natexlab{}.
\newblock \showarticletitle{{HBGG:} a Hierarchical Bayesian Geographical Model
  for Group Recommendation}. In \bibinfo{booktitle}{\emph{{SDM}}}.
  \bibinfo{pages}{372--380}.
\newblock


\bibitem[Mobasher et~al\mbox{.}(2007)]%
        {d3}
\bibfield{author}{\bibinfo{person}{Bamshad Mobasher}, \bibinfo{person}{Robin~D.
  Burke}, \bibinfo{person}{Runa Bhaumik}, {and} \bibinfo{person}{Chad
  Williams}.} \bibinfo{year}{2007}\natexlab{}.
\newblock \showarticletitle{Toward trustworthy recommender systems: An analysis
  of attack models and algorithm robustness}.
\newblock \bibinfo{journal}{\emph{{ACM} Trans. Internet Techn.}}
  \bibinfo{volume}{7}, \bibinfo{number}{4} (\bibinfo{year}{2007}),
  \bibinfo{pages}{23}.
\newblock


\bibitem[Moins et~al\mbox{.}(2020)]%
        {MoinsAB20}
\bibfield{author}{\bibinfo{person}{Th{\'{e}}o Moins}, \bibinfo{person}{Daniel
  Aloise}, {and} \bibinfo{person}{Simon~J. Blanchard}.}
  \bibinfo{year}{2020}\natexlab{}.
\newblock \showarticletitle{RecSeats: {A} Hybrid Convolutional Neural Network
  Choice Model for Seat Recommendations at Reserved Seating Venues}. In
  \bibinfo{booktitle}{\emph{RecSys}}. \bibinfo{pages}{309--317}.
\newblock


\bibitem[Morid et~al\mbox{.}(2014)]%
        {morid2014defending}
\bibfield{author}{\bibinfo{person}{Mohammad~Amin Morid}, \bibinfo{person}{Mehdi
  Shajari}, {and} \bibinfo{person}{Ali~Reza Hashemi}.}
  \bibinfo{year}{2014}\natexlab{}.
\newblock \showarticletitle{Defending recommender systems by influence
  analysis}.
\newblock \bibinfo{journal}{\emph{Inf. Retr.}} \bibinfo{volume}{17},
  \bibinfo{number}{2} (\bibinfo{year}{2014}), \bibinfo{pages}{137--152}.
\newblock


\bibitem[Pang et~al\mbox{.}(2018)]%
        {Pang0TZ18}
\bibfield{author}{\bibinfo{person}{Ming Pang}, \bibinfo{person}{Wei Gao},
  \bibinfo{person}{Min Tao}, {and} \bibinfo{person}{Zhi{-}Hua Zhou}.}
  \bibinfo{year}{2018}\natexlab{}.
\newblock \showarticletitle{Unorganized Malicious Attacks Detection}. In
  \bibinfo{booktitle}{\emph{NeurIPS}}. \bibinfo{pages}{6976--6985}.
\newblock


\bibitem[Sarwar et~al\mbox{.}(2001)]%
        {SarwarKKR01}
\bibfield{author}{\bibinfo{person}{Badrul~Munir Sarwar},
  \bibinfo{person}{George Karypis}, \bibinfo{person}{Joseph~A. Konstan}, {and}
  \bibinfo{person}{John Riedl}.} \bibinfo{year}{2001}\natexlab{}.
\newblock \showarticletitle{Item-based collaborative filtering recommendation
  algorithms}. In \bibinfo{booktitle}{\emph{{WWW}}}. \bibinfo{pages}{285--295}.
\newblock


\bibitem[Sedhain et~al\mbox{.}(2015)]%
        {SedhainMSX15}
\bibfield{author}{\bibinfo{person}{Suvash Sedhain},
  \bibinfo{person}{Aditya~Krishna Menon}, \bibinfo{person}{Scott Sanner}, {and}
  \bibinfo{person}{Lexing Xie}.} \bibinfo{year}{2015}\natexlab{}.
\newblock \showarticletitle{AutoRec: Autoencoders Meet Collaborative
  Filtering}. In \bibinfo{booktitle}{\emph{{WWW} (Companion Volume)}}.
  \bibinfo{pages}{111--112}.
\newblock


\bibitem[Seminario and Wilson(2014)]%
        {SeminarioW14b}
\bibfield{author}{\bibinfo{person}{Carlos~E. Seminario} {and}
  \bibinfo{person}{David~C. Wilson}.} \bibinfo{year}{2014}\natexlab{}.
\newblock \showarticletitle{Attacking item-based recommender systems with power
  items}. In \bibinfo{booktitle}{\emph{RecSys}}. \bibinfo{pages}{57--64}.
\newblock


\bibitem[Shi et~al\mbox{.}(2014)]%
        {ShiLH14}
\bibfield{author}{\bibinfo{person}{Yue Shi}, \bibinfo{person}{Martha Larson},
  {and} \bibinfo{person}{Alan Hanjalic}.} \bibinfo{year}{2014}\natexlab{}.
\newblock \showarticletitle{Collaborative Filtering beyond the User-Item
  Matrix: {A} Survey of the State of the Art and Future Challenges}.
\newblock \bibinfo{journal}{\emph{{ACM} Comput. Surv.}} \bibinfo{volume}{47},
  \bibinfo{number}{1} (\bibinfo{year}{2014}), \bibinfo{pages}{3:1--3:45}.
\newblock


\bibitem[Si and Li(2020)]%
        {si2020shilling}
\bibfield{author}{\bibinfo{person}{Mingdan Si} {and} \bibinfo{person}{Qingshan
  Li}.} \bibinfo{year}{2020}\natexlab{}.
\newblock \showarticletitle{Shilling attacks against collaborative recommender
  systems: a review}.
\newblock \bibinfo{journal}{\emph{Artif. Intell. Rev.}} \bibinfo{volume}{53},
  \bibinfo{number}{1} (\bibinfo{year}{2020}), \bibinfo{pages}{291--319}.
\newblock


\bibitem[Song et~al\mbox{.}(2020)]%
        {SongLHWLLG20}
\bibfield{author}{\bibinfo{person}{Junshuai Song}, \bibinfo{person}{Zhao Li},
  \bibinfo{person}{Zehong Hu}, \bibinfo{person}{Yucheng Wu},
  \bibinfo{person}{Zhenpeng Li}, \bibinfo{person}{Jian Li}, {and}
  \bibinfo{person}{Jun Gao}.} \bibinfo{year}{2020}\natexlab{}.
\newblock \showarticletitle{PoisonRec: An Adaptive Data Poisoning Framework for
  Attacking Black-box Recommender Systems}. In
  \bibinfo{booktitle}{\emph{{ICDE}}}. \bibinfo{pages}{157--168}.
\newblock


\bibitem[Su et~al\mbox{.}(2019)]%
        {su2019one}
\bibfield{author}{\bibinfo{person}{Jiawei Su},
  \bibinfo{person}{Danilo~Vasconcellos Vargas}, {and} \bibinfo{person}{Kouichi
  Sakurai}.} \bibinfo{year}{2019}\natexlab{}.
\newblock \showarticletitle{One Pixel Attack for Fooling Deep Neural Networks}.
\newblock \bibinfo{journal}{\emph{{IEEE} Trans. Evol. Comput.}}
  \bibinfo{volume}{23}, \bibinfo{number}{5} (\bibinfo{year}{2019}),
  \bibinfo{pages}{828--841}.
\newblock


\bibitem[Tang et~al\mbox{.}(2020)]%
        {revisit}
\bibfield{author}{\bibinfo{person}{Jiaxi Tang}, \bibinfo{person}{Hongyi Wen},
  {and} \bibinfo{person}{Ke Wang}.} \bibinfo{year}{2020}\natexlab{}.
\newblock \showarticletitle{Revisiting Adversarially Learned Injection Attacks
  Against Recommender Systems}. In \bibinfo{booktitle}{\emph{RecSys}}.
  \bibinfo{pages}{318--327}.
\newblock


\bibitem[Tao et~al\mbox{.}(2021)]%
        {single}
\bibfield{author}{\bibinfo{person}{Shuchang Tao}, \bibinfo{person}{Qi Cao},
  \bibinfo{person}{Huawei Shen}, \bibinfo{person}{Junjie Huang},
  \bibinfo{person}{Yunfan Wu}, {and} \bibinfo{person}{Xueqi Cheng}.}
  \bibinfo{year}{2021}\natexlab{}.
\newblock \showarticletitle{Single Node Injection Attack against Graph Neural
  Networks}. In \bibinfo{booktitle}{\emph{{CIKM}}}.
  \bibinfo{pages}{1794--1803}.
\newblock


\bibitem[van~der Maaten and Hinton(2008)]%
        {MaatenH08}
\bibfield{author}{\bibinfo{person}{Laurens van~der Maaten} {and}
  \bibinfo{person}{Geoffrey Hinton}.} \bibinfo{year}{2008}\natexlab{}.
\newblock \showarticletitle{Visualizing Data using t-SNE}.
\newblock \bibinfo{journal}{\emph{J. Mach. Learn. Res.}}  \bibinfo{volume}{9}
  (\bibinfo{year}{2008}), \bibinfo{pages}{2579--2605}.
\newblock


\bibitem[Wang et~al\mbox{.}(2018)]%
        {WangNL18}
\bibfield{author}{\bibinfo{person}{Cheng Wang}, \bibinfo{person}{Mathias
  Niepert}, {and} \bibinfo{person}{Hui Li}.} \bibinfo{year}{2018}\natexlab{}.
\newblock \showarticletitle{{LRMM:} Learning to Recommend with Missing
  Modalities}. In \bibinfo{booktitle}{\emph{{EMNLP}}}.
  \bibinfo{pages}{3360--3370}.
\newblock


\bibitem[Wang et~al\mbox{.}(2020)]%
        {WangNL2019}
\bibfield{author}{\bibinfo{person}{Cheng Wang}, \bibinfo{person}{Mathias
  Niepert}, {and} \bibinfo{person}{Hui Li}.} \bibinfo{year}{2020}\natexlab{}.
\newblock \showarticletitle{RecSys-DAN: Discriminative Adversarial Networks for
  Cross-Domain Recommender Systems}.
\newblock \bibinfo{journal}{\emph{{IEEE} Trans. Neural Networks Learn. Syst.}}
  \bibinfo{volume}{31}, \bibinfo{number}{8} (\bibinfo{year}{2020}),
  \bibinfo{pages}{2731--2740}.
\newblock


\bibitem[Wang et~al\mbox{.}(2019)]%
        {Wang0WFC19}
\bibfield{author}{\bibinfo{person}{Xiang Wang}, \bibinfo{person}{Xiangnan He},
  \bibinfo{person}{Meng Wang}, \bibinfo{person}{Fuli Feng}, {and}
  \bibinfo{person}{Tat{-}Seng Chua}.} \bibinfo{year}{2019}\natexlab{}.
\newblock \showarticletitle{Neural Graph Collaborative Filtering}. In
  \bibinfo{booktitle}{\emph{{SIGIR}}}. \bibinfo{pages}{165--174}.
\newblock


\bibitem[Williams et~al\mbox{.}(2007)]%
        {williams2007defending}
\bibfield{author}{\bibinfo{person}{Chad Williams}, \bibinfo{person}{Bamshad
  Mobasher}, {and} \bibinfo{person}{Robin~D. Burke}.}
  \bibinfo{year}{2007}\natexlab{}.
\newblock \showarticletitle{Defending recommender systems: detection of profile
  injection attacks}.
\newblock \bibinfo{journal}{\emph{Serv. Oriented Comput. Appl.}}
  \bibinfo{volume}{1}, \bibinfo{number}{3} (\bibinfo{year}{2007}),
  \bibinfo{pages}{157--170}.
\newblock


\bibitem[Wilson and Seminario(2013)]%
        {WilsonS13}
\bibfield{author}{\bibinfo{person}{David~C. Wilson} {and}
  \bibinfo{person}{Carlos~E. Seminario}.} \bibinfo{year}{2013}\natexlab{}.
\newblock \showarticletitle{When power users attack: assessing impacts in
  collaborative recommender systems}. In \bibinfo{booktitle}{\emph{RecSys}}.
  \bibinfo{pages}{427--430}.
\newblock


\bibitem[Wu et~al\mbox{.}(2021)]%
        {TrialAttack}
\bibfield{author}{\bibinfo{person}{Chenwang Wu}, \bibinfo{person}{Defu Lian},
  \bibinfo{person}{Yong Ge}, \bibinfo{person}{Zhihao Zhu}, {and}
  \bibinfo{person}{Enhong Chen}.} \bibinfo{year}{2021}\natexlab{}.
\newblock \showarticletitle{Triple Adversarial Learning for Influence based
  Poisoning Attack in Recommender Systems}. In
  \bibinfo{booktitle}{\emph{{KDD}}}. \bibinfo{pages}{1830--1840}.
\newblock


\bibitem[Wu et~al\mbox{.}(2023)]%
        {WuHWZW23}
\bibfield{author}{\bibinfo{person}{Le Wu}, \bibinfo{person}{Xiangnan He},
  \bibinfo{person}{Xiang Wang}, \bibinfo{person}{Kun Zhang}, {and}
  \bibinfo{person}{Meng Wang}.} \bibinfo{year}{2023}\natexlab{}.
\newblock \showarticletitle{A Survey on Accuracy-Oriented Neural
  Recommendation: From Collaborative Filtering to Information-Rich
  Recommendation}.
\newblock \bibinfo{journal}{\emph{{IEEE} Trans. Knowl. Data Eng.}}
  \bibinfo{volume}{35}, \bibinfo{number}{5} (\bibinfo{year}{2023}),
  \bibinfo{pages}{4425--4445}.
\newblock


\bibitem[Xing et~al\mbox{.}(2013)]%
        {XingMDSFL13}
\bibfield{author}{\bibinfo{person}{Xinyu Xing}, \bibinfo{person}{Wei Meng},
  \bibinfo{person}{Dan Doozan}, \bibinfo{person}{Alex~C. Snoeren},
  \bibinfo{person}{Nick Feamster}, {and} \bibinfo{person}{Wenke Lee}.}
  \bibinfo{year}{2013}\natexlab{}.
\newblock \showarticletitle{Take This Personally: Pollution Attacks on
  Personalized Services}. In \bibinfo{booktitle}{\emph{{USENIX} Security
  Symposium}}. \bibinfo{pages}{671--686}.
\newblock


\bibitem[Yang et~al\mbox{.}(2017)]%
        {covisit}
\bibfield{author}{\bibinfo{person}{Guolei Yang},
  \bibinfo{person}{Neil~Zhenqiang Gong}, {and} \bibinfo{person}{Ying Cai}.}
  \bibinfo{year}{2017}\natexlab{}.
\newblock \showarticletitle{Fake Co-visitation Injection Attacks to Recommender
  Systems}. In \bibinfo{booktitle}{\emph{{NDSS}}}.
\newblock


\bibitem[Yuan et~al\mbox{.}(2019)]%
        {YuanHZL19}
\bibfield{author}{\bibinfo{person}{Xiaoyong Yuan}, \bibinfo{person}{Pan He},
  \bibinfo{person}{Qile Zhu}, {and} \bibinfo{person}{Xiaolin Li}.}
  \bibinfo{year}{2019}\natexlab{}.
\newblock \showarticletitle{Adversarial Examples: Attacks and Defenses for Deep
  Learning}.
\newblock \bibinfo{journal}{\emph{{IEEE} Trans. Neural Networks Learn. Syst.}}
  \bibinfo{volume}{30}, \bibinfo{number}{9} (\bibinfo{year}{2019}),
  \bibinfo{pages}{2805--2824}.
\newblock


\bibitem[Yue et~al\mbox{.}(2021)]%
        {YueHZM21}
\bibfield{author}{\bibinfo{person}{Zhenrui Yue}, \bibinfo{person}{Zhankui He},
  \bibinfo{person}{Huimin Zeng}, {and} \bibinfo{person}{Julian~J. McAuley}.}
  \bibinfo{year}{2021}\natexlab{}.
\newblock \showarticletitle{Black-Box Attacks on Sequential Recommenders via
  Data-Free Model Extraction}. In \bibinfo{booktitle}{\emph{RecSys}}.
  \bibinfo{pages}{44--54}.
\newblock


\bibitem[Zeng et~al\mbox{.}(2023)]%
        {Zeng0JCL23}
\bibfield{author}{\bibinfo{person}{Meifang Zeng}, \bibinfo{person}{Ke Li},
  \bibinfo{person}{Bingchuan Jiang}, \bibinfo{person}{Liujuan Cao}, {and}
  \bibinfo{person}{Hui Li}.} \bibinfo{year}{2023}\natexlab{}.
\newblock \showarticletitle{Practical Cross-System Shilling Attacks with
  Limited Access to Data}. In \bibinfo{booktitle}{\emph{{AAAI}}}.
  \bibinfo{pages}{4864--4874}.
\newblock


\bibitem[Zhang et~al\mbox{.}(2020a)]%
        {ZhangLD020}
\bibfield{author}{\bibinfo{person}{Hengtong Zhang}, \bibinfo{person}{Yaliang
  Li}, \bibinfo{person}{Bolin Ding}, {and} \bibinfo{person}{Jing Gao}.}
  \bibinfo{year}{2020}\natexlab{a}.
\newblock \showarticletitle{Practical Data Poisoning Attack against Next-Item
  Recommendation}. In \bibinfo{booktitle}{\emph{{WWW}}}.
  \bibinfo{pages}{2458--2464}.
\newblock


\bibitem[Zhang et~al\mbox{.}(2021b)]%
        {ZhangTLSYZG21}
\bibfield{author}{\bibinfo{person}{Hengtong Zhang}, \bibinfo{person}{Changxin
  Tian}, \bibinfo{person}{Yaliang Li}, \bibinfo{person}{Lu Su},
  \bibinfo{person}{Nan Yang}, \bibinfo{person}{Wayne~Xin Zhao}, {and}
  \bibinfo{person}{Jing Gao}.} \bibinfo{year}{2021}\natexlab{b}.
\newblock \showarticletitle{Data Poisoning Attack against Recommender System
  Using Incomplete and Perturbed Data}. In \bibinfo{booktitle}{\emph{{KDD}}}.
  \bibinfo{pages}{2154--2164}.
\newblock


\bibitem[Zhang et~al\mbox{.}(2019)]%
        {ZhangYST19}
\bibfield{author}{\bibinfo{person}{Shuai Zhang}, \bibinfo{person}{Lina Yao},
  \bibinfo{person}{Aixin Sun}, {and} \bibinfo{person}{Yi Tay}.}
  \bibinfo{year}{2019}\natexlab{}.
\newblock \showarticletitle{Deep Learning Based Recommender System: {A} Survey
  and New Perspectives}.
\newblock \bibinfo{journal}{\emph{{ACM} Comput. Surv.}} \bibinfo{volume}{52},
  \bibinfo{number}{1} (\bibinfo{year}{2019}), \bibinfo{pages}{5:1--5:38}.
\newblock


\bibitem[Zhang et~al\mbox{.}(2020b)]%
        {Zhang20}
\bibfield{author}{\bibinfo{person}{Wei~Emma Zhang}, \bibinfo{person}{Quan~Z
  Sheng}, \bibinfo{person}{Ahoud Alhazmi}, {and} \bibinfo{person}{Chenliang
  Li}.} \bibinfo{year}{2020}\natexlab{b}.
\newblock \showarticletitle{Adversarial Attacks on Deep-learning Models in
  Natural Language Processing: A Survey}.
\newblock \bibinfo{journal}{\emph{ACM Trans. Intell. Syst. Technol.}}
  \bibinfo{volume}{11}, \bibinfo{number}{3} (\bibinfo{year}{2020}).
\newblock


\bibitem[Zhang et~al\mbox{.}(2021a)]%
        {ZhangCZWL21}
\bibfield{author}{\bibinfo{person}{Xuxin Zhang}, \bibinfo{person}{Jian Chen},
  \bibinfo{person}{Rui Zhang}, \bibinfo{person}{Chen Wang}, {and}
  \bibinfo{person}{Ling Liu}.} \bibinfo{year}{2021}\natexlab{a}.
\newblock \showarticletitle{Attacking Recommender Systems With Plausible
  Profile}.
\newblock \bibinfo{journal}{\emph{{IEEE} Trans. Inf. Forensics Secur.}}
  \bibinfo{volume}{16} (\bibinfo{year}{2021}), \bibinfo{pages}{4788--4800}.
\newblock


\bibitem[Zhang et~al\mbox{.}(2022)]%
        {zhang2022deep}
\bibfield{author}{\bibinfo{person}{Xin Zhang}, \bibinfo{person}{Zengmao Wang},
  {and} \bibinfo{person}{Bo Du}.} \bibinfo{year}{2022}\natexlab{}.
\newblock \showarticletitle{Deep Dynamic Interest Learning With Session Local
  and Global Consistency for Click-Through Rate Predictions}.
\newblock \bibinfo{journal}{\emph{{IEEE} Trans. Ind. Informatics}}
  \bibinfo{volume}{18}, \bibinfo{number}{5} (\bibinfo{year}{2022}),
  \bibinfo{pages}{3306--3315}.
\newblock


\bibitem[Zhang et~al\mbox{.}(2015)]%
        {zhang2015catch}
\bibfield{author}{\bibinfo{person}{Yongfeng Zhang}, \bibinfo{person}{Yunzhi
  Tan}, \bibinfo{person}{Min Zhang}, \bibinfo{person}{Yiqun Liu},
  \bibinfo{person}{Tat{-}Seng Chua}, {and} \bibinfo{person}{Shaoping Ma}.}
  \bibinfo{year}{2015}\natexlab{}.
\newblock \showarticletitle{Catch the Black Sheep: Unified Framework for
  Shilling Attack Detection Based on Fraudulent Action Propagation}. In
  \bibinfo{booktitle}{\emph{{IJCAI}}}. \bibinfo{pages}{2408--2414}.
\newblock


\bibitem[Zhou et~al\mbox{.}(2022)]%
        {abs-2202-13556}
\bibfield{author}{\bibinfo{person}{Kun Zhou}, \bibinfo{person}{Hui Yu},
  \bibinfo{person}{Wayne~Xin Zhao}, {and} \bibinfo{person}{Ji{-}Rong Wen}.}
  \bibinfo{year}{2022}\natexlab{}.
\newblock \showarticletitle{Filter-enhanced {MLP} is All You Need for
  Sequential Recommendation}.
\newblock \bibinfo{journal}{\emph{arXiv Preprint}} (\bibinfo{year}{2022}).
\newblock
\urldef\tempurl%
\url{https://arxiv.org/abs/2202.13556}
\showURL{%
\tempurl}


\end{thebibliography}

\end{document}